\documentclass[12pt,a4paper]{article}

\usepackage{amsmath,amssymb,a4,graphicx,color,ifpdf}
\usepackage[british]{babel}
\ifpdf\usepackage[pdftex]{hyperref}\else 
      \usepackage[dvips]{hyperref}
\fi

\definecolor{mathblue}{rgb}{0.2472, 0.24, 0.6}
\definecolor{mathred}{rgb}{0.6, 0.24, 0.442893}
\definecolor{mathyellow}{rgb}{0.6, 0.547014, 0.24}
\definecolor{mathgreen}{rgb}{0.24, 0.6, 0.33692}
\definecolor{palered}{rgb}{1,0.96,0.96}
\definecolor{palegreen}{rgb}{0.88,1,0.88}

\allowdisplaybreaks[4]

\newcommand{\di}{\genfrac{}{}{0pt}{}}
\def\under#1{\kern.4pt\underline{\kern-.4pt{}#1\kern-.4pt}\kern.4pt}

\renewcommand{\title}[1]{{\Large{\bf #1}\vspace{8mm}}}
\newcommand{\authors}[1]{\noindent{\large #1}\vspace{3mm}}
\newcommand{\address}[1]{{\itshape #1\vspace{2mm}}}

\makeatletter
\def\section{\@startsection{section}{1}{\z@}{-3.25ex plus -1ex minus
    -.2ex}{1.5ex plus .2ex}{\normalfont\large\bfseries}}
\def\subsection{\@startsection{subsection}{1}{\z@}{-3.25ex plus -1ex
    minus -.2ex}{1.5ex plus .2ex}{\normalfont\itshape}}

\def\@listI{\leftmargin\leftmargini
            \parsep 0.3ex  \@plus0.2ex \@minus 0.1ex
            \topsep 0.3ex  \@plus0.1ex  \@minus 0.1ex
            \itemsep 0.3ex  \@plus0.2ex \@minus 0.1ex}

\renewenvironment{thebibliography}[1]
         {\section*{References}\addcontentsline{toc}{section}{References}%
          \frenchspacing\small
          \begin{list}{[\arabic{enumi}]}
         {\usecounter{enumi}\parsep=2pt\@plus\p@\topsep 0pt
         \settowidth{\labelwidth}{[#1]}
         \leftmargin=\labelwidth\advance\leftmargin\labelsep
         \rightmargin=0pt\itemsep=0pt\sloppy}}{\end{list}}
 \makeatother

\newtheorem{Theorem}{Theorem}
\newtheorem{Proposition}[Theorem]{Proposition}
\newtheorem{Lemma}[Theorem]{Lemma}
\newtheorem{Definition}[Theorem]{Definition}

\begin{document}

\begin{center}

\title{Construction of the $\Phi^4_4$-quantum field theory on
  \\[0.5ex] 
noncommutative Moyal space$^*$}

\authors{Harald {\sc Grosse}$^1$ and Raimar {\sc Wulkenhaar}$^2$}

\address{$^{1}$\,Fakult\"at f\"ur Physik, Universit\"at Wien\\
Boltzmanngasse 5, A-1090 Wien, Austria}

\address{$^{2}$\,Mathematisches Institut der Westf\"alischen
  Wilhelms-Universit\"at\\
Einsteinstra\ss{}e 62, D-48149 M\"unster, Germany}

{\renewcommand\thefootnote{\fnsymbol{footnote}}
\footnotetext[1]{based on the lectures given at RIMS, 
Kyoto University, September 2013}}
\footnotetext[1]{harald.grosse@univie.ac.at}
\footnotetext[2]{raimar@math.uni-muenster.de}

\vskip 1cm

  {\bf Abstract} \\[1ex]
\begin{minipage}{13cm}
  We review our recent construction of the $\phi^4$-model on
  four-dimensional Moyal space. A milestone is the exact solution of
  the quartic matrix model $\mathcal{Z}[E,J]=\int d\Phi
  \exp(\mathrm{trace}(J\Phi- E\Phi^2 -\frac{\lambda}{4} \Phi^4))$ in
  terms of the solution of a non-linear equation for the 2-point
  function and the eigenvalues of $E$. The $\beta$-function vanishes
  identically. For the Moyal model, the theory of Carleman type
  singular integral equations reduces the construction to a fixed
  point problem. Its numerical solution reveals a second-order phase
  transition at $\lambda_c\approx -0.396$ and a phase transition of
  infinite order at $\lambda=0$.  The resulting Schwinger functions in
  position space are symmetric and invariant under the full Euclidean
  group. They are only sensitive to diagonal matrix correlation
  functions, and clustering is violated.  The Schwinger $2$-point
  function is reflection positive iff the diagonal matrix $2$-point
  function is a Stieltjes function.  Numerically this seems to be the
  case for coupling constants $\lambda \in [\lambda_c,0]$.
\end{minipage}
\end{center}

\section{Introduction}

Perturbatively renormalised quantum field theory is an enormous
phenomenological success, a success which lacks a mathematical
understanding. The perturbation series is at best an asymptotic
expansion which cannot converge at physical coupling constants. Some
physical effects such as confinement are out of reach for perturbation
theory. In two and partly three dimensions, methods of constructive
physics \cite{Glimm:1987ng, Rivasseau:1991ub}, often combined with the
Euclidean approach \cite{Schwinger:1959zz, Osterwalder:1973dx,
  Osterwalder:1974tc}, were used to rigorously establish quantum field
theory models.

In four dimensions there was little success so far. It is generally
believed that due to asymptotic freedom, non-Abelian gauge theory
(i.e.\ Yang-Mills theory) has the chance of a rigorous construction.
But this is a hard problem \cite{Jaffe:2000ne}.  What makes it so
difficult is the fact that any simpler model such as quantum
electrodynamics or the $\lambda\phi^4$-model cannot be constructed in
four dimensions (Landau ghost problem \cite{Landau:1954?a,
  Landau:1954?b, Landau:1954?c} or triviality \cite{Aizenman:1981du,
  Frohlich:1982tw}).

One of the main difficulties is the \emph{non-linearity} of the models
under consideration. Fixed point methods provide a standard approach
to non-linear problems, but they are rarely used in quantum field
theory. In this contribution we review a sequence of papers
\cite{Grosse:2012uv, Grosse:2013iva, Grosse:2014??} in which we
successfully used symmetry and fixed point methods to exactly solve a
toy model for a quantum field theory in four dimensions.

\begin{enumerate}
\item Following \cite{Grosse:2012uv}, we show in sec.~\ref{sec:qmm}
  that a Ward identity for the $U(\infty)$ group action leads to an
  exact solution of the quartic matrix model $\mathcal{Z}=\int
  \mathcal{D}[\Phi]\;
  \exp(\mathrm{trace}(J\Phi{-}E\Phi^2{-}\frac{\lambda}{4} \Phi^4))$ in
  terms of the solution of a non-linear equation.  As by-product we
  find that any renormalisable quartic matrix model has vanishing
  $\beta$-function. All these steps are completely elementary.

\item Self-dual $\phi^4_4$-theory on Moyal space \cite{Grosse:2004yu,
    Grosse:2004ik} is of that type. For extreme noncommutativity
  $\theta\to \infty$, and after careful discussion of thermodynamic
  and continuum limit, the non-linear equation is reduced to a
  fixed-point problem \cite{Grosse:2012uv} which has a unique
  non-perturbative and non-trivial solution for $\lambda < 0$
  \cite{Grosse:2014??}.  Sec.~\ref{sec:Moyal} reviews this work.  The
  key step is the observation that a certain difference function
  satisfies a linear singular integral equation of Carleman type
  \cite{Carleman, Tricomi}. We also present some numerical results,
  contained in work in progress \cite{Grosse:2014??}, which show
  evidence for phase transitions.

\item Following \cite{Grosse:2013iva}, we identify in
  sec.~\ref{sec:Schwinger} a limit to Schwinger functions for a scalar
  field on $\mathbb{R}^4$. Surprisingly for a highly noncommutative
  model, these Schwinger functions show full Euclidean symmetry.
  Otherwise they have unusual properties such as absent momentum
  transfer in interaction processes. This seems to suggest triviality,
  but the numerical investigation \cite{Grosse:2014??} of the 2-point
  function shows scattering remnants from a noncommutative geometrical
  substructure.  Most surprisingly, the Schwinger 2-point function
  seems to be reflection positive in one of its phases.

\end{enumerate}

\section{Exact solution of the quartic matrix model}
\label{sec:qmm}

For us a `matrix' is a compact (Hilbert-Schmidt) operator on Hilbert
space $H=L^2(I,\mu)$. Such operators $\Phi\in \mathcal{L}^2(H)$ can be
represented by integral kernel operators $(\Phi v)_a=\int_I d\mu_b\,
\Phi_{ab} v_b$. Then all natural matrix operations such as product,
adjoint and trace have counterparts $(\Phi \Phi')_{ab}=\int_I d\mu_c\;
\Phi_{ac}\Phi'_{cb}$, $(\Phi^*)_{ab}=\overline{\Phi_{ba}}$ and
$\mathrm{tr}(\Phi\Phi')=\int_I d\mu_a \; (\Phi \Phi')_{aa}$ in
$\mathcal{L}^2(H)$.

To define a \emph{Euclidean quantum field theory} for a matrix $\Phi\in
\mathcal{L}^2(H)$ we give ourselves an action functional 
\begin{equation}
S[\Phi]= V\,\mathrm{tr}(E \Phi^2 + P[\Phi])\;.
\label{action}
\end{equation}
Here, $P[\Phi]$ is a polynomial in $\Phi$ with scalar coefficients,
and this alone would be a familiar action in the theory of matrix
models \cite{Di Francesco:1993nw}. To be closer to field theory on a
(compact) manifold $\mathcal{M}$ we add the analogue of the kinetic
term $\int_{\mathcal{M}} dx\; (-\Delta \phi) \phi$, that is, we
require the external matrix $E$ to be an unbounded selfadjoint
positive operator on $H$ with compact resolvent. The volume $V$ will
play a crucial r\^ole. The construction involves several
regularisation and limiting procedures. One such regularisation
consists in a finite size $\mathcal{N}$ for the matrices, and $V$ will
be a certain function of $\mathcal{N}$ which together with
$\mathcal{N}$ is sent to $\infty$.

Adding a source term to the action, we define the \emph{partition
  function} as 
\begin{align}
\mathcal{Z}[J]=\int \mathcal{D}[\Phi]
\;\exp(-S[\Phi]+V\,\mathrm{tr}(\Phi J))\;,
\end{align}
where $\mathcal{D}[\Phi]$ is the extension of the Lebesgue measure
from finite-rank operators to $\mathcal{L}^2(H)$ and $J$ a test
function matrix. For absent $P[\Phi]\mapsto 0$ in (\ref{action}),
$\mathcal{D}[\Phi]\exp(-V\,\mathrm{tr}(E\Phi^2))/\mathcal{Z}[0]$ 
would be the Gau\ss{}ian
measure of covariance determined by $E$. What we want, and what we
achieve, is to construct (the moments of) the measure
$\mathcal{D}[\Phi]\exp(-V\,\mathrm{tr}(E\Phi^2
+\frac{\lambda}{4}\Phi^4))/\mathcal{Z}[0]$
in the limit $V\to \infty$. Such a
limit cannot be expected for $\mathcal{Z}$. Instead, we pass to the
generating functional $\log \mathcal{Z}[J]$ of \emph{connected
  correlation functions},
\begin{align}
\langle \varphi_{a_1b_1}\dots\varphi_{a_Nb_N}\rangle_c = \frac{\partial^N
 \log \mathcal{Z}[J]}{ \partial J_{b_1a_1}\dots \partial J_{b_Na_N}}\Big|_{J=0}\;.
\end{align}

\subsection{Ward identity and topological expansion}

Unitary operators $U$ belonging to an appropriate unitisation of the
compact operators on $H$ give rise to a transformation $\Phi \mapsto
\tilde{\Phi}=U\Phi U^*$. Since the space of selfadjoint compact
operators is invariant under the adjoint action, we have
\[
\int \mathcal{D}[\Phi]
\;\exp(-S[\Phi]+V\,\mathrm{tr}(\Phi J))
=
\int \mathcal{D}[\tilde{\Phi}]
\;\exp(-S[\tilde{\Phi}]+V\,\mathrm{tr}(\tilde{\Phi} J))\;.
\]
Unitary invariance $\mathcal{D}[\tilde{\Phi}]=
\mathcal{D}[\Phi]$ of the Lebesgue measure implies 
\[
0=\int \mathcal{D}[\Phi
\;\Big\{\exp(-S[\Phi]+V\,\mathrm{tr}(\Phi J))
-\exp(-S[\tilde{\Phi}]+V\,\mathrm{tr}(\tilde{\Phi} J))\Big\}\;.
\]
Note that the integrand $\{\dots\}$ itself does not vanish
because $\mathrm{tr}(E \Phi^2)$ and $\mathrm{tr}(\Phi J)$ are not
unitarily invariant; we only have 
$\mathrm{tr}(P[\Phi]){=}\mathrm{tr}(P[\tilde{\Phi}])$ due to
$UU^*{=}U^*U{=}\mathrm{id}$ together with the trace property.
Linearisation of $U$ about the identity operator 
leads to the \emph{Ward identity} 
\begin{align}
0=\int \mathcal{D}[\Phi]
\;\Big\{E\Phi\Phi-\Phi\Phi E - J\Phi + \Phi J\Big\}
\exp(-S[\Phi]+V\,\mathrm{tr}(\Phi J))\;.
\label{Ward-pre}
\end{align}
We can always place ourselves in an orthonormal basis of $H$ where $E$
is diagonal (but $J$ is not). Since $E$ is of compact resolvent, $E$
has eigenvalues $E_a>0$ of finite multiplicity $\mu_a$.  We thus label
the matrices by an enumeration of the (necessarily discrete)
eigenvalues of $E$ and an enumeration of the basis vectors of the
finite-dimensional eigenspaces.  Writing $\Phi$ in $\{\dots\}$ of
(\ref{Ward-pre}) as functional derivative $\Phi_{ab}=
\frac{\partial}{V \partial J_{ba}}$, we have proved 
(first obtained in \cite{Disertori:2006nq}):
\begin{Proposition}
  The partition function $\mathcal{Z}[J]$ of the matrix model defined
  by the external matrix $E$ satisfies the $|I|\times |I|$ Ward
  identities
\begin{align}
0=\sum_{n\in I}\Big(
\frac{(E_a-E_p)}{V} \frac{\partial^2\mathcal{Z}}{\partial J_{an}\partial J_{np}}
+J_{pn} \frac{\partial\mathcal{Z}}{\partial J_{an}}
-J_{na} \frac{\partial\mathcal{Z}}{\partial J_{np}}
\Big)\;.
\label{Ward}
\end{align}
\end{Proposition}
Without loss of generality we can assume that the map $I \ni m \mapsto
E_m \in \mathbb{R}_+$ is injective. Namely, correlation functions
will only depend on the set of eigenvalues $(E_m)$ of $E$.
Partitioning the index set $I$ into equivalence classes $[m]$ which
have the same $E_m$, the index sum over a function that only depends
on $E_m$ becomes $\sum_{m\in I} f(m)= \sum_{[m]\in [I]} \mu_{[m]}
f([m])$.  Therefore, at the expense of adding a measure
$\mu_{[m]}=\mathrm{dim} \ker (E-E_m\mathrm{id})$, we can assume that $m
\mapsto E_m$ is injective.

In a perturbative expansion, Feynman graphs in matrix models 
are \emph{ribbon graphs}. Viewed as
simplicial complexes, they encode the topology $(B,g)$ of
a genus-$g$ Riemann surface with $B$ boundary components
(or punctures, marked points, holes, broken/external faces). 
Some simple examples for $P[\Phi]=\Phi^4$ are:
\begin{align*}
\begin{array}{r} B=1 \\ g=1
\end{array}
&
\begin{picture}(60,24)
\put(0,0){\includegraphics[width=4.2cm]{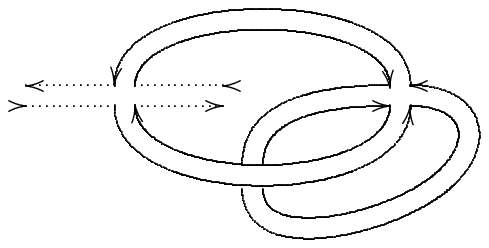}}
\put(6,15.5){\mbox{\scriptsize$a$}}
\put(6,10.5){\mbox{\scriptsize$b$}}
\put(13,15.5){\mbox{\scriptsize$a$}}
\put(13,10.5){\mbox{\scriptsize$b$}}
\end{picture} &
\begin{array}{r} B=2 \\ g=0
\end{array}&
\begin{picture}(40,29)
\put(0,-3.5){\includegraphics[width=3.5cm]{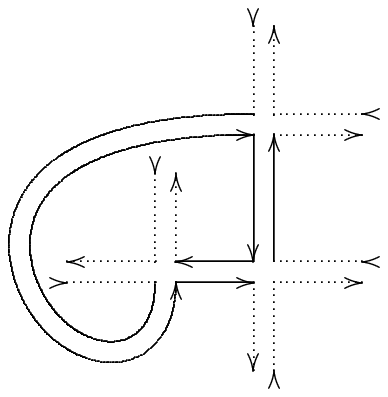}}
\put(12.5,12.5){\mbox{\scriptsize$p$}}
\put(17.5,12.5){\mbox{\scriptsize$q$}}
\put(11,8.5){\mbox{\scriptsize$p$}}
\put(11,3.5){\mbox{\scriptsize$q$}}
\put(22,-0.5){\mbox{\scriptsize$a$}}
\put(27,-0.5){\mbox{\scriptsize$b$}}
\put(30,3.5){\mbox{\scriptsize$b$}}
\put(30,8.5){\mbox{\scriptsize$c$}}
\put(30,17.5){\mbox{\scriptsize$c$}}
\put(30,22.5){\mbox{\scriptsize$d$}}
\put(27,24.5){\mbox{\scriptsize$d$}}
\put(22,24.5){\mbox{\scriptsize$a$}}
\end{picture}
\end{align*}
Since $E$ is diagonal, the matrix index is conserved along each strand
of the ribbon graph. We have to distinguish between internal faces
(with constant matrix index) and broken faces which constitute
the boundary components. Such a boundary face is characterised by $N_\beta\geq
1$ external double lines to which we attach the source matrices $J$.
Conservation of the matrix index along each strand implies that the right
index of $J_{ab}$ coincides with the left index of another $J_{bc}$,
or of the same $J_{bb}$. Accordingly, the $\beta^{\mathrm{th}}$ boundary
component carries a cycle $J^{N_\beta}_{p_1\dots
  p_{N_\beta}}:=\prod_{j=1}^{N_\beta} J_{p_jp_{j+1}}$ of $N_\beta$ external
source matrices, with $N_\beta+1\equiv 1$.

Being interested in a non-perturbative solution, we will not expand
the partition function into ribbon graphs. But we keep the topological
information and expand $\log \mathcal{Z}[J]$ according to the cycle
structure:
\begin{align}
\log\frac{ \mathcal{Z}[J]}{\mathcal{Z}[0]} 
=\sum_{B=1}^\infty \sum_{1\leq N_1 \leq \dots \leq
  N_B}^\infty 
\sum_{p_1^\beta,\dots,p^\beta_{N_\beta} \in I}
\frac{V^{2-B} }{S_{N_1\dots N_B}} 
G_{|p_1^1\dots p_{N_1}^1|\dots|p_1^B\dots p^B_{N_B}|} 
\prod_{\beta=1}^B \Big(\frac{J^{N_\beta}_{p_1^\beta\dots
    p^\beta_{N_\beta}}}{N_\beta}\Big) \;.
\label{logZ}
\end{align}
The symmetry factor $S_{N_1\dots N_B}$ is obtained as follows: 
If $\nu_i$ of the $B$ numbers $N_\beta$ in a given tuple 
$(N_1,\dots,N_B)$ are equal to $i$, then 
$S_{N_1\dots N_B}=\prod_{i=1}^{N_B} \nu_i!$.

Next we turn the Ward identity (\ref{Ward}) for injective $m\mapsto
E_m$ into a formula for the second derivative $\sum_{n\in I} \frac{\partial^2
  \mathcal{Z}[J]}{\partial J_{an} \partial J_{np} }$ of 
the partition function.  The $J$-cycle
structure in $\log \mathcal{Z}$ creates

\begin{itemize}
\item singular contributions $\sim \delta_{ap}$,

\item regular contributions present for all $a,p$:
\end{itemize}

\begin{Theorem}
\begin{align}
\sum_{n\in I} 
\frac{\partial^2 \mathcal{Z}[J]}{\partial J_{an} \partial J_{np} }
&= 
\delta_{ap} \Big\{
V^2\sum_{(K)} \frac{J_{P_1}\cdots J_{P_K}}{S_{(K)}}
\Big(\sum_{n\in I}  \frac{G_{|an|P_1|\dots |P_K|}}{V^{|K|+1}}
{+} \frac{G_{|a|a|P_1|\dots |P_K|}}{V^{|K|+2}}
\nonumber
\\
&\hspace*{1.5cm}
+\sum_{r\geq 1} \sum_{q_1,\dots, q_r \in I} 
\frac{G_{|q_1a q_1\dots q_r| P_1|\dots|P_K|} J^{r}_{q_1\dots q_r}}{
V^{|K|+1}} \Big) 
\nonumber
\\
& \quad + V^4\!\!\!
\sum_{(K),(K')} \!\!\! \frac{J_{P_1}{\cdots} J_{P_K} J_{Q_1}
{\cdots} J_{Q_{K'}}}{S_{(K)} S_{(K')}}
 \frac{G_{|a| P_1|\dots |P_K|}}{V^{|K|+1}} 
\frac{G_{|a|Q_1|\dots |Q_{K'}|}}{V^{|K'|+1}} \Big\} \mathcal{Z}[J]
\nonumber
\\
&+ \frac{V}{E_{p}-E_{a}}\sum_{n \in I}\Big(
J_{pn}\frac{\partial \mathcal{Z}[J]}{\partial J_{an}}
{-}J_{na}\frac{\partial\mathcal{Z}[J]}{\partial J_{np}}\Big)\;.
\label{ZJJ}
\end{align}

\end{Theorem}
\emph{Proof.} We identify the following four sources 
of a singular contribution $\sim \delta_{ap}$:
\begin{enumerate} 
\item 
$\displaystyle
\sum_n \frac{\partial^2}{\partial J_{an} \partial J_{np}} 
\sum_{q_1,q_2,\dots} G_{\dots|q_1q_2|\dots}
\Big(\frac{\stackrel{\downarrow}{J_{q_1q_2}}
\stackrel{\downarrow}{J_{q_2q_1}}}{2}\Big) \prod J
$

\item 
$\displaystyle
\sum_n \frac{\partial^2}{\partial J_{an} \partial J_{np}} 
\sum_{q_1,q_2,\dots} G_{\dots|q_1|\dots|q_2|\dots}
\Big(\frac{\stackrel{\downarrow}{J_{q_1q_1}}}{1}\Big) 
\Big(\frac{\stackrel{\downarrow}{J_{q_2q_2}}}{1}\Big) 
\prod J
$

\item 
$\displaystyle
\sum_n \frac{\partial}{\partial J_{an}} 
\frac{\partial}{\partial J_{np}} 
\sum_{q_0,\dots,q_{r+1},\dots} \!\!\!\!\!\!\!\! 
G_{\dots|q_0 q_1\dots q_r q_{r+1}|\dots}
\Big(\frac{J_{q_0q_1}J_{q_1q_2}{\cdots}
  J_{q_rq_{r+1}} \!\!\! \stackrel{\downarrow}{J_{q_{r+1}q_0}}}{r+2}\Big)
\prod J\hspace*{-3cm}
$
\\
$\displaystyle
= \sum_n \frac{\partial}{\partial J_{an}} 
\sum_{q_1,\dots,q_r,\dots} \!\!G_{\dots|p q_1\dots q_r n|\dots}
\Big(\stackrel{\downarrow}{J_{pq_1}}J_{q_1q_2}\cdots  J_{q_r n} \Big)
\prod J
$

\item 
$\displaystyle
\sum_n \frac{\partial^2}{\partial J_{an} \partial J_{np}} 
\Big[\sum_{q_1,\dots} G_{\dots|q_1|\dots}
\Big(\frac{\stackrel{\downarrow}{J_{q_1q_1}}}{1}\Big) \prod J
\Big]
\Big[\sum_{q_2,\dots} G_{\dots|q_2|\dots}
\Big(\frac{\stackrel{\downarrow}{J_{q_2q_2}}}{1}\Big) \prod J
\Big]\hspace*{-3cm}
$

\end{enumerate}
All other types of derivatives, collected into $\big(\sum_{n\in I}
\frac{\partial^2 \mathcal{Z}[J]}{\partial J_{an} \partial J_{np}
}\big)_{\mathrm{reg}}$, persist for $a\neq p$. For $p\neq a$ we clearly
have
\begin{align}
\Big(\sum_{n\in I} 
\frac{\partial^2 \mathcal{Z}[J]}{\partial J_{an} \partial J_{np}}
\Big)_{\mathrm{reg}}=\sum_{n\in I} 
\frac{\partial^2 \mathcal{Z}[J]}{\partial J_{an} \partial J_{np}}
\Big|_{a\neq p}
= \frac{V}{E_p-E_a} \Big(J_{pn} \frac{\partial\mathcal{Z}}{\partial J_{an}}
-J_{na} \frac{\partial\mathcal{Z}}{\partial J_{np}}\Big)\;,
\label{Ward-reg}
\end{align}
where the last equality is the Ward identity (\ref{Ward}), divided by
$\frac{E_p-E_a}{V}\neq 0$. By a continuity argument, the rightmost
term in (\ref{Ward-reg}) must agree with $\big(\sum_{n\in I} 
\frac{\partial^2 \mathcal{Z}[J]}{\partial J_{an} \partial J_{np}
}\big)_{\mathrm{reg}}$ also in the limit $p\to a$, and this finishes
the proof. \hfill $\square$%

\subsection{Schwinger-Dyson equations}

We can write the action as 
$S=\frac{V}{2} \sum_{a,b} (E_a+E_b) \Phi_{ab} \Phi_{ba} 
+ V S_{int}[\Phi]$, where $E_a$ are the eigenvalues of $E$.
Functional integration yields, up to an irrelevant constant,
\begin{align}
\mathcal{Z}[J]=e^{-V S_{int}[\frac{\partial}{V \partial J}]}
e^{\frac{V}{2}\langle J,J\rangle_E}\;,\qquad
\langle J,J\rangle_E:=\sum_{m,n\in I} \frac{J_{mn}J_{nm}}{E_m+E_n}\;.
\label{calZ}
\end{align}
Instead of a perturbative expansion of $e^{-V
  S_{int}[\frac{\partial}{V \partial J}]}$, we apply those
$J$-derivatives to (\ref{calZ}) which give rise to a correlation
function $G_{\dots}$ on the lhs. On the rhs of (\ref{calZ}), these
external derivatives combine with internal derivatives from
$S_{int}[\frac{\partial}{V \partial J}]$ to certain identities for
$G_{\dots}$.  These Schwinger-Dyson equations are often of little use
because they express an $N$-point function in terms of $(N{+}2)$-point
functions.

In the field-theoretical matrix models under consideration, 
the Ward identity (\ref{ZJJ}) lets this tower of Schwinger-Dyson
equations collapse. To see this we consider the 2-point function
$G_{|ab|}$ for $a\neq b$. According to (\ref{logZ}), $G_{|ab|}$ is obtained by 
deriving (\ref{calZ}) with respect to $J_{ba}$ and $J_{ab}$:  
\begin{align}
G_{|ab|} &= \! \frac{1}{V \mathcal{Z}[0]}
\frac{\partial^2\mathcal{Z}[J] }{\partial J_{ba}\partial J_{ab}}
\Big|_{J=0}
\qquad\qquad\parbox[c]{4cm}{\scriptsize
(disconnected part of $\mathcal{Z}$ does not contribute for $a\neq b$)}
\nonumber
\\
& =
\frac{1}{V\mathcal{Z}[0]} \Big\{
{\frac{\partial}{\partial J_{ba}}}
e^{-V S_{int}\big[\tfrac{\partial}{V\partial J}\big]}
{\frac{\partial}{\partial J_{ab}}}
e^{\frac{V}{2} \langle J,J\rangle_E}
\Big\}_{J=0}
\nonumber
\\
& =
\frac{1}{(E_{a}+E_b)\mathcal{Z}[0]} \Big\{
{\frac{\partial}{\partial J_{ba}}}
e^{-V S_{int}\big[\tfrac{\partial}{V\partial J}\big]}
{J_{ba}}
e^{\frac{V}{2} \langle J,J\rangle_E}
\Big\}_{J=0}
\nonumber
\\
&= \! {\frac{{1}}{E_a+E_b}} +
\frac{1}{{(E_a+E_b)}\mathcal{Z}[0]} \Big\{
\Big( {\Phi_{ab}}{\frac{\partial(-V S_{int})}{
\partial \Phi_{ab}}}\Big)
\Big[\frac{\partial}{V \partial J}\Big] \Big\}\mathcal{Z}[J]\Big|_{J=0}\;.
\label{SD-1}
\end{align}
Now observe that 
$\frac{\partial(-VS_{int})}{\partial \Phi_{ab}}$ contains,
for any $P[\Phi]$, the derivative
$\sum_{n}\frac{  \partial^2}{
\partial J_{an} \partial J_{np} }$ which we know from (\ref{ZJJ}).
In case of the quartic matrix model 
$P[\Phi]=\frac{\lambda}{4}\Phi^4$ we have 
$\frac{\partial(-V S_{int})}{\partial \Phi_{ab}}
= -\lambda V \sum_{n,p\in I} \Phi_{bp}\Phi_{pn}\Phi_{na}$, hence 
\[\Big( {\Phi_{ab}}{\frac{\partial(-V S_{int})}{
\partial \Phi_{ab}}}\Big)
\Big[\frac{\partial}{V \partial J}\Big]
= -\frac{\lambda}{V^3} \sum_{p,n\in I}
\frac{\partial^2}{\partial J_{pb} {\partial J_{ba} }}\frac{\partial^2}{
\partial J_{an} \partial J_{np}} \;,
\]
and the Schwinger-Dyson equation (\ref{SD-1}) for $G_{|ab|}$ becomes
with (\ref{ZJJ})
\begin{align}
G_{|ab|}
&=   \frac{1}{E_a+E_b} -
\frac{\lambda}{{V^3} (E_a+E_b)\mathcal{Z}[0]} 
\sum_{p\in I} \frac{\partial^2}{
\partial J_{pb} \partial J_{ba}}
{\sum_{n\in I} \frac{\partial^2 \mathcal{Z}}{
\partial J_{an} \partial J_{np}}}\Big|_{J=0}
\nonumber
\\
&=
 \frac{1}{E_a+E_b} -
\frac{\lambda}{{V}(E_a+E_b)\mathcal{Z}[0]} 
\frac{\partial^2}{
\partial J_{ab} \partial J_{ba}}
\Big\{
\nonumber
\\
&\quad
\Big(\sum_{n\in I} \frac{G_{|an|}}{V}
+\sum_{n,q,r\in I} \frac{G_{|an|qr|}}{V^2} \frac{J_{qr}J_{rq}}{2}
+
\sum_{n,q,r\in I} \frac{G_{|an|q|r|}}{V^3} \frac{J_{qq}}{1} 
\frac{J_{rr}}{1}
\nonumber
\\
&\quad  
+ \frac{G_{|a|a|}}{V^2}
+\sum_{q,r\in I} \frac{G_{|a|a|qr|}}{V^3} \frac{J_{qr}J_{rq}}{2}
+\sum_{q,r\in I} \frac{G_{|a|a|q|r|}}{V^4} \frac{J_{qq}}{1} 
\frac{J_{rr}}{1}
\nonumber
\\
& \quad + \sum_{q,r \in I} 
\frac{G_{|qa qr|}}{V} J_{qr}J_{rq}
+ V^2
 \frac{G_{|a| q|}}{V^2} \frac{J_{qq}}{1}
 \frac{G_{|a| r|}}{V^2} \frac{J_{rr}}{1}\Big)
\mathcal{Z}[J]\Big\}
\bigg|_{J=0}
\nonumber
\\
& -
\frac{\lambda}{V^2 (E_a+E_b)\mathcal{Z}[0]}
\sum_{p\in I} \frac{
\Big(
\frac{\partial^2 \mathcal{Z}[J]}{\partial J_{ab} \partial  J_{ba}}
+\delta_{pb}\frac{ \partial^2 \mathcal{Z}[J]}{\partial J_{aa} \partial  J_{bb}}
-\frac{\partial^2 \mathcal{Z}[J]}{\partial J_{pb} \partial  J_{bp}}
\Big)}{E_{p}-E_{a}}\Bigg|_{J=0}\;.
\end{align}
Taking 
$\frac{\partial^2 \mathcal{Z}[J]}{\partial J_{pb} \partial  J_{bp}}
= (V G_{|pb|}+ \delta_{pb} G_{|p|b|})\mathcal{Z}[0] +\mathcal{O}(J)$ and 
$\frac{\partial J_{rr}}{\partial J_{ab}}=0$ for $a\neq b$ into
account, we have proved:
\begin{Proposition} 
The $2$-point function of a quartic matrix model with action
$S=V\,\mathrm{tr}(E\Phi^2+\frac{\lambda}{4}\Phi^4)$ satisfies for
injective $m\mapsto E_m$ the Schwinger-Dyson equation
\begin{subequations}
\label{Gab-allg}
\begin{align}
G_{|ab|}
&= \frac{1}{E_a+E_b}
- \frac{\lambda}{E_a+E_b} \frac{1}{V}\sum_{p\in I}
\Big( G_{|ab|} G_{|ap|}
- \frac{G_{|pb|}-G_{|ab|}}{E_p-E_a}\Big) &&\left. \rule{0mm}{6mm} \right\}
\label{Gab-g+0}
\\
& \begin{array}[c]{@{}l@{}}
\displaystyle 
- \frac{\lambda}{V^2(E_a+E_b)}\Big( 
G_{|a|a|}G_{|ab|}
+ \frac{1}{V}\sum_{n\in I} G_{|an|ab|}
\\
\qquad\qquad\qquad\qquad\displaystyle  
+ G_{|aaab|}+ G_{|baba|}
-\frac{G_{|b|b|}-G_{|a|b|}}{E_b-E_a}\Big)
\end{array}
&& \left. \rule{0mm}{10mm} \right\}
\label{Gab-g+1}
\\
&
- \frac{\lambda}{V^4(E_a+E_b)} G_{|a|a|ab|}\;. 
&&\left. \rule{0mm}{6mm} \right\}
\label{Gab-g+2}
\end{align}
\end{subequations}

\end{Proposition}
It can be checked \cite{Grosse:2012uv} that in a genus expansion
$G_{\dots} = \sum_{g=0}^\infty V^{-2g} G^{(g)}_{\dots}$ (which is
probably not convergent but Borel summable), precisely the line
(\ref{Gab-g+0}) preserves the genus, the lines (\ref{Gab-g+1})
increase $g\mapsto g+1$ and the line (\ref{Gab-g+2}) increases
$g\mapsto g+2$.  In particular, in a scaling limit $V\to \infty$ with
$\frac{1}{V}\sum_{p\in I}$ finite, the exact Schwinger-Dyson equation
for $G_{|ab|}$ coincides with its restriction (\ref{Gab-g+0}) to the 
planar sector $g=0$, a closed non-linear equation for $G^{(0)}_{|ab|}$
alone:
\begin{align}
G^{(0)}_{|ab|}
= \frac{1}{E_a+E_b}
- \frac{\lambda}{E_a+E_b} \frac{1}{V}\sum_{p\in I}
\Big( G^{(0)}_{|ab|} G^{(0)}_{|ap|}
- \frac{G^{(0)}_{|pb|}-G^{(0)}_{|ab|}}{E_p-E_a}\Big)\;.
\label{Gab}
\end{align}
We have derived in 2007/08 this self-consistency equation for the Moyal
model by the graphical method proposed by \cite{Disertori:2006nq}.  In
this form, (\ref{Gab}) is meaningless because $\sum_{p\in I}$
diverges.  In 2009 we solved the renormalisation problem, namely the
renormalisation of infinitely many Feynman graphs at once
\cite{Grosse:2009pa}. This renormalisation increases the
non-linearity.  In \cite{Grosse:2009pa} we have solved (\ref{Gab})
perturbatively to $\mathcal{O}(\lambda^3)$.  After several years of
setbacks with the non-perturbative solution, a breakthrough came in
2012: The equation (\ref{Gab}) can be turned into an equation which is
linear in the difference $G^{(0)}_{|ab|}-G^{(0)}_{|a0|}$ to the
boundary and non-linear only in $G^{(0)}_{|a0|}$!

A similar calculation gives the Schwinger-Dyson equation for higher
$N$-point functions:
\begin{subequations}
\label{GN-all}
\begin{align}
&G_{|ab_1\dots b_{N-1}|}
\nonumber
\\[-0.5ex]
& 
\begin{array}[c]{@{}l@{}} 
\displaystyle
= - \frac{\lambda}{E_a+E_{b_1}} \bigg(
\frac{1}{V}\sum_{p\in I} \Big(
 G_{|ap|} G_{|ab_1\dots b_{N-1}|} 
- \frac{G_{|pb_1\dots b_{N-1}|}-G_{|ab_1\dots b_{N-1}|}}{E_p-E_a}\Big)
\\
\qquad\qquad\quad 
\displaystyle
-\sum_{l=1}^{\frac{N-2}{2}} G_{|b_1\dots b_{2l}|}
\frac{
  G_{|b_{2l+1}\dots b_{N-1}a|}- G_{|b_{2l+1}\dots
    b_{N-1}b_{2l}|}}{E_{b_{2l}}-E_a}
\bigg)
\end{array} && \left. \rule{0mm}{14mm} \right\}
\label{GN-g+0}
\\
& \begin{array}[c]{@{}l@{}}
\displaystyle 
- \frac{\lambda}{V^2(E_a+E_{b_1})}\Big( 
G_{|a|a|}G_{|ab_1\dots b_{N-1}|}
+ \sum_{k=1}^{N-1} G_{|b_1\dots b_ka b_k\dots b_{N-1}a|}
\\
\displaystyle 
\qquad\qquad\quad + G_{|aaab_1\dots b_{N-1}|}
+ \frac{1}{V}\sum_{n\in I} G_{|an|ab_1\dots b_{N-1}|}
\\[-0.5ex]
\qquad\qquad\quad
\displaystyle
-\sum_{k=1}^{N-1} \frac{G_{|b_1\dots b_k|b_{k+1}\dots 
b_{N-1}b_k|}-G_{|b_1\dots b_k|b_{k+1}\dots b_{N-1} a|}}{E_{b_k}-E_a}\Big)
\end{array}
&& \left. \rule{0mm}{17mm} \right\}
\label{GN-g+1}
\\
&
- \frac{\lambda}{V^4(E_a+E_{b_1})} G_{|a|a|ab_1\dots b_{N-1}|}\;.
&& \left. \rule{0mm}{6mm} \right\}
\label{GN-g+2}
\end{align}
\end{subequations}
Again, the first lines (\ref{GN-g+0}) preserve the genus, whereas
$g\mapsto g+1$ in (\ref{GN-g+1}) and $g\mapsto g+2$ in (\ref{GN-g+2}).
The planar sector $G^{(0)}_{|ab_1\dots b_{N-1}|}$, exact for $V\to
\infty$ with $\frac{1}{V}\sum_{p\in I}$ finite, is a linear
inhomogeneous equation with inductively known parameters.

It turns out that a real theory with $\Phi=\Phi^*$ admits a short-cut
which directly gives the higher $N$-point functions without any index
summation. Since the equations for $G_{\dots}$ are real and
$\overline{J_{ab}}= J_{ba}$, the reality
$\mathcal{Z}=\overline{\mathcal{Z}}$ implies (in addition to
invariance under cyclic permutations) invariance under orientation
reversal
\begin{align}
G_{|p^1_0p^1_1\dots p^1_{N_1-1}|\dots
|p^B_0p^B_1\dots p^B_{N_B-1}|}
=
G_{|p^1_0p^1_{N_1-1}\dots p^1_{1}|\dots
|p^B_0p^B_{N_B-1}\dots p^B_{1}|}\;.
\label{reality}
\end{align}
Whereas empty for $G_{|ab|}$, in $(E_a{+}E_{b_1}) G_{ab_1b_2\dots
  b_{N-1}} - (E_a{+}E_{b_{N-1}}) G_{ab_{N-1}\dots b_2 b_1}$ the
identities (\ref{reality}) lead to many cancellations which result in
a universal algebraic recursion formula:
\begin{Proposition}
\begin{align}
G_{|b_0b_1\dots b_{N-1}|}
&=(-\lambda)
\sum_{l=1}^{\frac{N-2}{2}}
\frac{G_{|b_0 b_1 \dots b_{2l-1}|} G_{|b_{2l}b_{2l+1}\dots b_{N-1}|}
- G_{|b_{2l} b_1 \dots b_{2l-1}|} G_{|b_0 b_{2l+1}\dots b_{N-1}|}
}{(E_{b_0}-E_{b_{2l}})(E_{b_1}-E_{b_{N-1}})}
\nonumber
\\
& + \frac{(-\lambda)}{V^2}
\sum_{k=1}^{N-1}
\frac{G_{|b_0 b_1 \dots b_{k-1}|b_{k}b_{k+1}\dots b_{N-1}|}
- G_{|b_{k} b_1 \dots b_{k-1}|b_0 b_{k+1}\dots b_{N-1}|}
}{(E_{b_0}-E_{b_{k}})(E_{b_1}-E_{b_{N-1}})}\;.
\label{GN-real}
\end{align}
\end{Proposition}
The last line of (\ref{GN-real}) increases the genus and is 
absent in $G^{(0)}_{|b_0b_1\dots b_{N-1}|}$. Instead of giving the general
proof, let us look at the case $N=4$. Then (\ref{GN-all}), multiplied by
$E_{a}-E_{b_1}$, reads
\begin{align}
&(E_a-E_b)G_{|abcd|}
\nonumber
\\
&= (- \lambda) \bigg(
\frac{1}{V}\sum_{p\in I} \Big(
 G_{|ap|} G_{|abcd|} 
- \frac{G_{|pbcd|}-G_{|abcd|}}{E_p-E_a}\Big)
- G_{|bc|}
\frac{G_{|da|}- G_{|dc|}}{E_c-E_a}\bigg)
\nonumber
\\
& 
- \frac{\lambda}{V^2}\Big( 
G_{|a|a|}G_{|abcd|}
+ G_{|babcda|}+G_{|bcacda|}+G_{|bcdada|}
+G_{|aaabcd|}
+ \frac{1}{V}\sum_{p\in I} G_{|ap|abcd|}
\nonumber
\\
& \qquad\qquad\quad
- \frac{G_{|b|cdb|}-G_{|b|cda|}}{E_b-E_a}
- \frac{G_{|bc|dc|}-G_{|bc|da|}}{E_c-E_a}
- \frac{G_{|d|bcd|}-G_{|a|bcd|}}{E_d-E_a}\Big)
\nonumber
\\
&
- \frac{\lambda}{V^4} G_{|a|a|abcd|}\;.
\label{G4}
\end{align}
Write down the same equation but with $b\leftrightarrow d$, and take
the difference between these equations. Then most terms cancel because
by (\ref{reality}) we have the equalities $G_{|abcd|}=G_{|adcb|}$, 
$G_{|pbcd|}=G_{|pdcb|}$, 
$G_{|babcda|}= G_{|dcbaba|}$, $G_{|bcacda|}=
G_{|dcacba|}$, $G_{|bcdada|}=G_{|dadcba|}$,
$G_{|aaabcd|}=G_{|aaadcb|}$, $G_{|ap|abcd|}=
G_{|ap|adcb|}$, $G_{|b|cdb|}=G_{|dcb|b|}$,
$G_{|bc|dc|}=G_{|dc|bc|}$, 
$G_{|bcd|d|}=G_{|d|cbd|}$ 
and $G_{|a|a|abcd|}=G_{|a|a|adcb|}$. Altogether, the difference 
(\ref{G4})$-$(\ref{G4})$_{b\leftrightarrow d}$ reads after cancellation
\begin{align*}
(E_d-E_b) G_{|abcd|} 
&= (-\lambda) \frac{G_{|ab|}G_{|cd|}-G_{|ad|}G_{|cb|}}{E_c-E_a}
\\
&-\frac{\lambda}{V^2} \Big(
\frac{G_{|b|cda|}-G_{|a|cdb|}}{E_b-E_a}
+\frac{G_{|bc|da|}-G_{|ba|dc|}}{E_c-E_a}
+\frac{G_{|a|bcd|}-G_{|d|bca|}}{E_d-E_a}\Big)\;,
\end{align*}
and this is (\ref{GN-real}) for $N=4$. 

For completeness, we list in the appendix the Schwinger-Dyson equation 
for $B=2$ boundary components. 

We make the following key observation: An affine transformation $E\mapsto 
ZE+C$ together with a corresponding rescaling $\lambda\mapsto Z^2\lambda$
leaves the algebraic equations (\ref{GN-real}) as well as 
(\ref{GN-B=2-odd}) and (\ref{GN-B=2-even}) invariant:
\begin{Theorem}
  Given a real quartic matrix model with 
$S=V\,\mathrm{tr}(E\Phi^2+\frac{\lambda}{4} \Phi^4)$ and $m\mapsto
  E_m$ injective, which determines the set 
$G_{|p^1_1 \dots p^1_{N_1}|\dots |p^B_1\dots p^B_{N_B}|}$ of
($N_1{+}\dots {+} N_B$)-point functions. 
Assume that the basic functions with all $N_i\leq 2$ 
are turned finite by $E_a \mapsto
  Z(E_a+\frac{\mu^2}{2}-\frac{\mu_{bare}^2}{2})$ and $\lambda
  \mapsto Z^2\lambda$. 
Then all functions with one $N_i\geq 3$

\begin{enumerate}
\item are finite without further need of a renormalisation of
  $\lambda$, i.e.\ all renormalisable quartic matrix models have
  vanishing $\beta$-function,

\item are given by universal algebraic recursion formulae in terms of
  renormalised basic functions with $N_i\leq 2$. \hfill $\square$%
\end{enumerate}
\end{Theorem}
The theorem tells us that vanishing of the $\beta$-function for the
self-dual $\Phi^4_4$-model on Moyal space (proved in
\cite{Disertori:2006nq} to all orders in perturbation theory) is
generic to all quartic matrix models, and the result even holds
non-perturbatively!

The universal recursion formula (\ref{GN-real}) computes the planar
$N$-point function $G_{|b_0\dots b_{N-1}|}$ at $B=1$ as a sum of
fractions with products of $2$-point functions in the numerator and
products of differences of eigenvalues of $E$ in the denominator. This
structure admits an 
interesting graphical interpretation. We draw the indices $b_0,\dots
b_{N-1}$ in cyclic order on the circle $S^1$ and represent a factor
$G_{b_ib_j}$ as a chord connecting $b_i$ with $b_j$ and a factor 
$\frac{1}{E_{b_i}-E_{b_j}}$ as an arrow from $b_i$ to $b_j$: 
\begin{align}
G^{(0)}_{|b_0b_1b_2b_3|}
&=(-\lambda)\frac{G_{b_0b_1}G_{b_2b_3}{-}
G_{b_0b_3}G_{b_2b_1}}{(E_{b_0}-E_{b_2})(E_{b_1}-E_{b_3})}
=  (-\lambda) 
\left\{
\includegraphics[viewport=0 25 60 60]{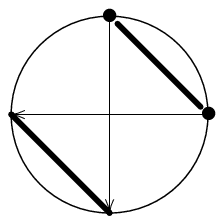}
+ 
\includegraphics[viewport=0 25 60 60]{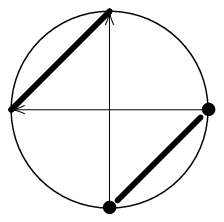}
\right\}\;,
\nonumber
\\
G_{b_0\dots b_5} 
&= (-\lambda)^2 \left\{
\includegraphics[viewport=0 25 60 60]{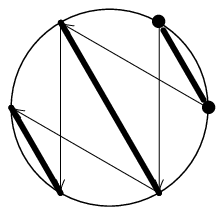}
+ 
\includegraphics[viewport=0 25 60 60]{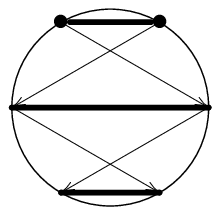}
+ 
\includegraphics[viewport=0 25 60 60]{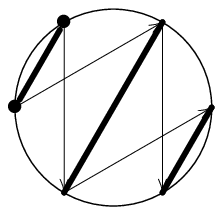}
\right.
\nonumber
\\
&  \qquad\qquad +
\left( 
\includegraphics[viewport=0 25 60 60]{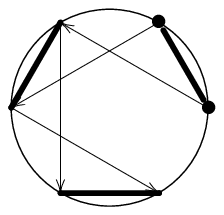}
+ 
\includegraphics[viewport=0 25 60 60]{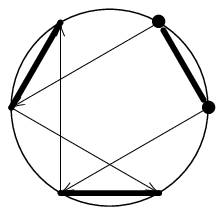}
+ 
\includegraphics[viewport=0 25 60 60]{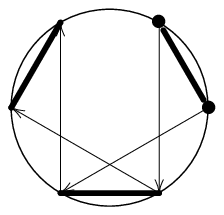}
\right) 
\nonumber
\\
& \qquad\qquad + 
\left. \left( 
\includegraphics[viewport=0 25 60 60]{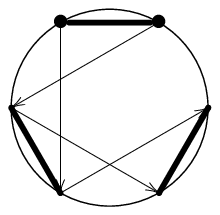}
+ 
\includegraphics[viewport=0 25 60 60]{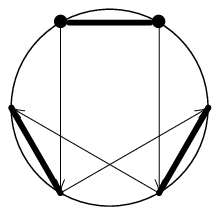}
+ 
\includegraphics[viewport=0 25 60 60]{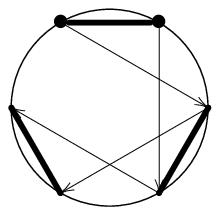}
\right) \right\}\;.
\end{align}
The chords form the non-crossing chord diagrams counted by the Catalan
number $C_{\frac{N}{2}}{=}\frac{N!}{(\frac{N}{2}{+}1)!\frac{N}{2}!}$.
The arrows form two disjoint trees, one connecting the even vertices
and one connecting the odd vertices.  By rational fraction expansion
it is possible to achieve that each tree intersects the chord only in
the vertices.  The assignment of trees to a given chord diagram is, in
general, not unique. A canonical choice is not known to us.

\subsection{Digression: Quantum gravity in two dimensions}

Two-dimensional quantum gravity (see \cite{Di Francesco:1993nw,
  Ambjorn:1997di} for reviews) can be interpreted as the enumeration
of random triangulations of surfaces. Its asymptotic behaviour is
captured by the matrix model partition function
\begin{align}
\mathcal{Z}=\int \mathcal{D}[\Phi] \; \exp\Big (- \mathcal{N} \sum_n t_n\;
\mathrm{tr}(\Phi^n)\Big)\;,
\end{align}
where the integral is over ($\mathcal{N}\times \mathcal{N}$)-Hermitean
matrices $\Phi$ and the $t_n$ are scalar coefficients. In the limit
$\mathcal{N}\to \infty$, this series in $(t_n)$ is evaluated in terms
of the $\tau$-function for the Korteweg-de Vries (KdV) hierarchy.
There is another approach to topological gravity in which the
partition function is a series in $(t_n)$ with coefficients given by
intersection numbers of complex curves.  Witten conjectured
\cite{Witten:1990hr} that the partition functions of the two
approaches coincide. This conjecture was proved by Kontsevich
\cite{Kontsevich:1992ti} who achieved the computation of the
intersection numbers in terms of weighted sums over ribbon graphs (fat
Feynman graphs), which he proved to be generated from the Airy
function matrix model (Kontsevich model)
\begin{align}
\mathcal{Z}[E]=
\frac{\displaystyle \int \mathcal{D}[\Phi] \;
\exp \big(-\tfrac{1}{2}\mathrm{tr}(E\Phi^2)
+\tfrac{\mathrm{i}}{6} \mathrm{tr}(\Phi^3)\big)}{
\displaystyle \int \mathcal{D}[\Phi] \;\exp 
\big(-\tfrac{1}{2}\mathrm{tr}(E\Phi^2)\big)}\;.
\label{Kontsevich}
\end{align}
The external matrix $E=E^*>0$ is related 
by $t_n
=(2n{-}1)!!\;
\mathrm{tr}(E^{-(2n-1)})$ to the series $(t_n)$. 
The limit $\mathcal{N} \to \infty$ of $\mathcal{Z}[E]$
gives the KdV evolution equation, thus proving
Witten's conjecture.

We have proved that also the \emph{quartic matrix model}
\begin{align}
\mathcal{Z}[E,J,\lambda]=
\frac{\displaystyle
\int \mathcal{D}[\Phi] \;\exp \big(-\mathrm{tr}(E\Phi^2)
+\mathrm{tr}(J\Phi)
-\tfrac{\lambda}{4} \mathrm{tr}(\Phi^4)\big)
}{\displaystyle
\int \mathcal{D}[\Phi] \;\exp \big(-\mathrm{tr}(E\Phi^2)
-\tfrac{\lambda}{4} \mathrm{tr}(\Phi^4)\big)}
\label{quartic}
\end{align}
is in the large-$\mathcal{N}$ limit exactly solvable in terms of the
solution of a non-linear equation (\ref{Gab}). Any triangulation can be
subdivided into a quadrangulation 
\[
\parbox{5cm}{\unitlength 0.5mm\begin{picture}(40,16)
\put(0,0){\line(1,0){60}}
\put(0,0){\line(2,1){40}}
\put(60,0){\line(-1,1){20}}
\put(30,8){\line(-1,-1){8}}
\put(30,8){\line(6,1){19}}
\put(30,8){\line(-3,1){9}}
\end{picture}}
\]
(and vice versa). From Witten's uniqueness argument
\cite{Witten:1990hr}, 2D quantum gravity should have equivalent
descriptions as cubic (\ref{Kontsevich}) and quartic (\ref{quartic})
matrix model. Understanding the precise relation between 
(\ref{Kontsevich}) and (\ref{quartic}) would be of high interest:
\begin{enumerate}
\item In contrast to (\ref{quartic}), the cubic action
(\ref{Kontsevich}) lacks manifest positivity due to its purely
imaginary coupling constant. 

\item A quartic action admits a Hubbard-Stratonovich transform which
  is the key ingredient of a new approach to constructive quantum
  field theory \cite{Rivasseau:2007fr} that avoids the cluster
  expansion.

\item Conversely, the integrability of (\ref{Kontsevich}) might
  provide valuable information about the solution of the
  self-consistency equation (\ref{Gab}).

\end{enumerate}

\smallskip

Coloured tensor models (see \cite{Gurau:2011xp, Rivasseau:2013uca} for
recent reviews) extend these methods to quantum gravity in $D\geq 3$.
They became a very active domain of research after understanding
\cite{Gurau:2010ba} of the analogue of the large-$\mathcal{N}$
behaviour of matrix models \cite{'tHooft:1973jz}.  They have
Schwinger-Dyson equations (see e.g.\ \cite{Bonzom:2012cu}) and action
of the $U(\infty)$ group. A first promising result is the recent
derivation of closed equations for the 2-point functions of rank 3 and
4 tensorial group field theory \cite{Samary:2014tja}.

\section{$\Phi^4_4$-theory on Moyal space as a fixed point problem}
\label{sec:Moyal}

\subsection{Preliminaries}

Taking the renormalisation group \cite{Wilson:1973jj} serious, we
would expect that General Relativity, because not renormalisable, is
irrelevant and hence scaled away. The existence of gravity thus tells
us that the scaling must stop at some length scale, and from the
weakness of the gravitational coupling constant one deduces the value
of that scale: the Planck length $10^{-35}\,\mathrm{m}$. There, the
geometry of nature is expected to differ from the familiar structure
of a differentiable manifold. One of many candidates for Planck scale
physics is noncommutative geometry \cite{Connes:1994yd}, a vast
reformulation of geometry and topology in the language of operator
algebras. The focus is shifted from manifolds to generalisations of
the algebra of functions. This concept proved very successful in
understanding the geometry of the Standard Model of particle physics
as Riemannian geometry of a space which is the product of a manifold
with a discrete space \cite{Connes:1996gi, Chamseddine:1996zu}.

A large class of examples of noncommutative geometries comes from
deformations of the algebra of functions on manifolds. Schwartz
functions on Euclidean space $\mathbb{R}^4$ admit an
$\mathbb{R}^4$-group action by translation. As shown by Rieffel
\cite{Rieffel:1993}, this group action induces a noncommutative 
associative product on the space
of Schwartz functions, the
Moyal product:
\begin{align}
(f\star g)(x)= \int_{\mathbb{R}^4 \times \mathbb{R}^4} \frac{dy\;dk}{(2\pi)^4}
f(x{+}\tfrac{1}{2}\Theta k) \; g(x{+}y)\; e^{\mathrm{i} \langle
  k,y\rangle}\;,\quad
\Theta=-\Theta^t\in M_4(\mathbb{R})\;. 
\end{align}

Whether or not the Moyal space $(\mathbb{R}^4,\star)$ is relevant for
Planck scale physics is pure speculation (although a refinement can be
justified by uncertainty relations for position operators
\cite{Doplicher:1994tu}).  In any case the Moyal space is a nice toy
model on which it is easy to formulate and to study (quantum) field
theories. To formulate a Euclidean quantum field theory on Moyal space
it is, at first sight, enough to replace in the action of a usual
field theory the pointwise product of functions by the
$\star$-product. The simplest example is the $\phi^{\star 4}_4$-model with
action
\begin{align}
S[\phi]=\int_{\mathbb{R}^4} dx\; \Big(\frac{1}{2}
\phi \star (-\Delta+ \mu^2)\phi +\frac{\lambda}{4} 
\phi\star\phi\star\phi\star \phi\Big)(x)\;.
\label{UVIR}
\end{align}
The resulting Feynman rules \cite{Filk:1996dm} lead to situations
where a multiple insertion of non-planar subgraphs gives rise to
divergences of arbitrarily high degree (ultraviolet/infrared mixing
\cite{Minwalla:1999px}). See \cite{Chepelev:2000hm} for a thorough
investigation of this problem. Relativistic quantum field theories on
noncommutative Minkowski space are much more difficult
\cite{Bahns:2002vm}. Here the UV/IR-mixing problem occurs in different
types of graphs \cite{Bahns:2010dx}.

The Moyal algebra $(\mathcal{S}(\mathbb{R}^4),\star)$ has a matrix
basis \cite{GraciaBondia:1987kw, Varilly:1988jk, Gayral:2003dm}
\begin{align}
\phi(x)&=\sum_{\under{m},\under{n}\in \mathbb{N}^2}
\Phi_{\under{m}\under{n}} f_{\under{m}\under{n}}(x),\quad
\qquad 
f_{\under{m}\under{n}}(x)=f_{m_1n_1}(x^0,x^1)f_{m_2n_2}(x^3,x^4)\;,
\nonumber
\\
f_{mn}(y^0,y^1)&=
2 (-1)^m \sqrt{\frac{m!}{n!}}\Big(\sqrt{\frac{2}{\theta}}
y\Big)^{n-m} L^{n-m}_m\Big(\frac{2|y|^2}{\theta}\Big)
e^{-\frac{|y|^2}{\theta}}\;,
\label{matrixbasis}
\end{align}
where $L^n_m$ are Laguerre polynomials, $y \equiv y^0{+}\mathrm{i}y^1$
and $\under{m}=(m_1,m_2)$. Without loss of generality we assume the
only non-vanishing components of $\Theta$ to be
$\theta:=\Theta_{12}={-}\Theta_{21}=\Theta_{34}={-}\Theta_{43}$.  The
functions $f_{\under{m}\under{n}}$ satisfy
\[
(f_{\under{k}\under{l}}{\star} f_{\under{m}\under{n}})(x)
=\delta_{\under{m}\under{l}}
f_{\under{k}\under{n}}(x)\;, \qquad \int_{\mathbb{R}^4}dx
\;f_{\under{m}\under{n}}(x)=(2\pi\theta)^2
\delta_{\under{m}\under{n}}\;.
\]
Therefore, the $\phi^{\star 4}_4$-interaction in (\ref{UVIR}) becomes
a matrix product (we write $\phi$ for a function and $\Phi$ for a matrix):
\begin{align}
S[\phi] =
(2\pi\theta)^2 \sum_{\under{k},\under{l},\under{m},
\under{n}\in \mathbb{N}^2} \!\!\!\!\Big(
\tfrac{1}{2} \Phi_{\under{k}\under{l}} (\Delta_{\under{k}\under{l};
\under{m}\under{n}}+ \mu^2\delta_{\under{k}\under{n}}
\delta_{\under{l}\under{m}})\Phi_{\under{m}\under{n}} 
+\frac{\lambda}{4} \Phi_{\under{k}\under{l}}\Phi_{\under{l}\under{m}} 
\Phi_{\under{m}\under{n}} \Phi_{\under{n}\under{k}}  \Big)\;.
\label{GW-matrix}
\end{align}
The matrix kernel $\Delta_{\under{k}\under{l};
\under{m}\under{n}}$ of the Laplacian $(-\Delta)$, viewed as map from 
$\mathbb{N}^4$ to $\mathbb{N}^4$, consists of a local interaction 
plus nearest neighbour interaction.

In \cite{Grosse:2004yu} we studied the renormalisation group flow of
the $\phi^{\star 4}_4$-model in matrix representation (using a
power-counting theorem \cite{Grosse:2003aj} for matrix models with
kernel $\Delta_{\under{k}\under{l}; \under{m}\under{n}}$).  We noticed
that the marginal parts of the local term and of the nearest neighbour
term in $\Delta_{\under{k}\under{l}; \under{m}\under{n}}$ have
\emph{different flows}. To absorb these different flows a 4$^\text{th}$
relevant/marginal operator in the action functional is necessary. This
operator corresponds to a harmonic oscillator potential:
\begin{equation}
S[\phi]= 64\pi^2 \!\!\!\int \!\!d^4x
\Big( \dfrac{Z}{2} \phi {\star} \big({-}\Delta {+}
\Omega^2 (2\Theta^{-1}x)^2 + \mu_{bare}^2\big) \phi
+ \frac{\lambda Z^2}{4} \phi {\star}
\phi {\star} \phi {\star} \phi\Big)(x)\;.
\label{GW}
\end{equation}
We proved in \cite{Grosse:2004yu} that the corresponding Euclidean
quantum field theory is renormalisable to all orders in perturbation
theory. This result was reestablished by various methods, see
\cite{Rivasseau:2007ab} for a review.

Presence of the harmonic oscillator term $\Omega\neq 0$ breaks
translation invariance. Conversely, this term achieves covariance
under Langmann-Szabo duality transformation \cite{Langmann:2002cc}
which consists in exchanging $x\leftrightarrow p$ and $\phi(x)
\leftrightarrow \hat{\phi}(p)$ followed by Fourier transform back to
the original variables. Remarkably, this transformation leaves
$\displaystyle \int dx\;\phi{\star}\phi{\star}\phi{\star} \phi$
invariant, and it exchanges $\int dx\;\phi (-\Delta)\phi$ with $\int
dx\; \phi |2\Theta^{-1}x|^2\phi$. Presence of the oscillator term
gives rise to an interesting spectral noncommutative geometry
\cite{Gayral:2011vu} (see also \cite{Grosse:2007jy}) which is
conceptually simpler than the isospectral deformation
\cite{Gayral:2003dm} of $\mathbb{R}^4$. Most importantly, the
oscillator term cures the Landau ghost problem \cite{Landau:1954?a,
  Landau:1954?b, Landau:1954?c} of usual $\phi^4_4$-theory: We have
discovered in \cite{Grosse:2004by, Grosse:2004ik} that the one-loop
renormalisation group flows of $\Omega$ and $\lambda$ influence each
other in such a way that the running coupling constant
$\lambda(\Lambda)$ remains finite at any scale $\Lambda$. Even more,
at the self-duality point $\Omega=1$ the $\beta$-function of the
$\lambda\Phi^4_4$-coupling vanishes to all orders in perturbation
theory \cite{Disertori:2006nq}.  This result was obtained by an
ingenious combination of Ward identities and Schwinger-Dyson equations
(see \cite{Disertori:2006uy} for an explicit three-loop calculation).
In \cite{Grosse:2012uv} we have generalised the method of
Disertori-Gurau-Magnen-Rivasseau \cite{Disertori:2006nq} to the whole
class of quartic matrix models (reviewed in sec.~\ref{sec:qmm}).
Vanishing of the $\beta$-function is often connected with
integrability, and together with the absent Landau ghost problem a
non-perturbatively constructed $\phi^4_4$-model on Moyal space came
into reach.  The first milestone was the derivation of the
self-consistency equation (\ref{Gab}) and the understanding of its
renormalisation in \cite{Grosse:2009pa}.  It took us several years to
fully understand this equation, and it is only recently that we
finished the solution/construction of the Moyal space $\phi^4_4$-model
\cite{Grosse:2012uv}. In the sequel we review this construction.

\subsection{Renormalisation and integral representation}
\label{sec:renorm}

At the self-duality point $\Omega=1$, the matrix kernel
$\Delta_{\under{k}\under{l}; \under{m}\under{n}}^{\Omega=1}$ of the
Schr\"odinger operator $H=-\Delta+ \|2\Theta^{-1}x\|^2$ becomes purely
local and turns the action (\ref{GW}) in matrix basis
(\ref{matrixbasis}) into a (field-theoretical) quartic matrix model
with action
\begin{align}
S[\Phi]&=  V \bigg( 
\sum_{\under{m},\under{n} \in \mathbb{N}^2_{\mathcal{N}} }\!\!\!
E_{\under{m}}\, \Phi_{\under{m}\under{n}} \Phi_{\under{n}\under{m}} 
+\frac{Z^2\lambda}{4} \!\!\!\!
\sum_{\under{m},\under{n},\under{k},\under{l} 
\in \mathbb{N}^2_{\mathcal{N}} } \!\!\!\!\!\!\!
\Phi_{\under{m}\under{n}} \Phi_{\under{n}\under{k}} 
\Phi_{\under{k}\under{l}} \Phi_{\under{l}\under{m}}\bigg)\;,
\label{Sphi}
\\
E_{\under{m}}
& = Z\Big( \frac{|\under{m}|}{\sqrt{V}}
+\frac{\mu_{bare}^2}{2} \Big) \;,\qquad
|\under{m}|:=m_1+m_2 \leq \mathcal{N}\;,~~
V=\Big(\frac{\theta}{4}\Big)^2\;.\nonumber
\end{align}
Our general results on quartic matrix models imply that the planar 
$2$-point function $G^{(0)}_{|\under{a}\under{b}|}$ 
satisfies the self-consistency equation (\ref{Gab}),
\begin{align}
G_{|\under{a}\under{b}|}^{(0)}
= \frac{1}{E_{\under{a}}+E_{\under{b}}}
- \frac{Z^2 \lambda}{E_{\under{a}}+E_{\under{b}}} 
\frac{1}{V}\sum_{\under{p}\in \mathbb{N}^2_{\mathcal{N}}}
\Big( G^{(0)}_{|\under{a}\under{b}|} G^{(0)}_{|\under{a}\under{p}|}
- \frac{G^{(0)}_{|\under{p}\under{b}|}-G^{(0)}_{|\under{a}\under{b}|}}{
E_{\under{p}}-E_{\under{a}}}\Big)\;.
\label{Gab-bare}
\end{align}
We have introduced a cut-off $\mathbb{N}^2_{\mathcal{N}}$ in the
matrix size; the index sum diverges for
$\mathbb{N}^2_{\mathcal{N}}\mapsto \mathbb{N}^2$. As usual, the
renormalisation strategy consists in adjusting $Z,\mu_{bare}$ in such a
way that the limit $\mathbb{N}^2_{\mathcal{N}}\mapsto \mathbb{N}^2$
exists. This will be achieved by normalisation conditions for the 
1PI function $\Gamma_{\under{a}\under{b}}$ defined by
$G^{(0)}_{|\under{a}\under{b}|}=: 
(H_{\under{a}\under{b}}-\Gamma_{\under{a}\under{b}})^{-1}$, where
$H_{\under{a}\under{b}}:=E_{\under{a}}+E_{\under{b}}$. 
We express (\ref{Gab-bare}) in terms of
$\Gamma_{\under{a}\under{b}}$, 
\begin{align}
\Gamma_{\under{a}\under{b}}&=  
-\frac{\lambda Z^2}{V} \!\!  \sum_{\under{p} \in
\mathbb{N}^2_{\mathcal{N}} }  \Big(\frac{1}{H_{\under{a}\under{p}}
-\Gamma_{\under{a}\under{p}}}
+\frac{1}{H_{\under{p}\under{b}}-\Gamma_{\under{p}\under{b}}}
- \frac{1}{(H_{\under{p}\under{b}}-\Gamma_{\under{p}\under{b}})} 
\frac{ \Gamma_{\under{p}\under{b}} -\Gamma_{\under{a}\under{b}}}{
\frac{Z}{\sqrt{V}} (|\under{p}|{-}|\under{a}|)}\Big)\;,
\label{Gammaab}
\end{align}
and write $\Gamma_{\under{a}\under{b}}$ as first-order Taylor formula
with remainder $\Gamma_{\under{a}\under{b}}^{ren}$,
\[
\Gamma_{\under{a}\under{b}}= Z \mu_{bare}^2 -\mu^2 
+ \tfrac{(Z-1)}{\sqrt{V}} (|\under{a}|+|\under{b}|) 
+\Gamma^{ren}_{\under{a}\under{b}}\;, \quad
\Gamma^{ren}_{\under{0}\under{0}}=0\;, \quad 
(\partial \Gamma^{ren})_{\under{0}\under{0}}=0\;.
\]
Equation (\ref{Gammaab}) for $\Gamma_{\under{a}\under{b}}\big[
\Gamma^{ren}_{\under{a}\under{b}},\mu_{bare}^2,Z\big]$ together with
$\Gamma^{ren}_{\under{0}\under{0}}=0$ and $(\partial
\Gamma^{ren})_{\under{0}\under{0}}$ constitute three equations to
determine the three functions
$\Gamma_{\under{a}\under{b}}^{ren},\mu_{bare}^2,Z$. Eliminating
$\mu_{bare}^2,Z$ thus gives rise to a \emph{closed equation for 
the renormalised function $\Gamma^{ren}_{\under{a}\under{b}}$ alone}.
For this elimination it is important to note that the equations for
$\Gamma_{\under{a}\under{b}}^{ren},\mu_{bare}^2,Z$ depend on
$\under{a},\under{b}$ only via the norms $|\under{a}|,|\under{b}|$ which
parametrise the spectrum of $E$.  Therefore,
$\Gamma_{\under{a}\under{b}}$ is actually a function only of
$|\under{a}|,|\under{b}|$, and consequently the index sum reduces to
$\sum_{\under{p} \in \mathbb{N}^2_{\mathcal{N}}} f(|\under{p}|) =
\sum_{|\under{p}|=0}^{\mathcal{N}} (|\under{p}|{+}1)f(|\under{p}|)$.

We study a particular scaling limit in which matrix size
$\mathcal{N}$ and volume $V$ are simultaneously sent to $\infty$ 
such that the ratio $\frac{\mathcal{N}}{\sqrt{V \mu^4}}=
\Lambda^2 (1{+}\mathcal{Y})$ is kept fixed. Note that
$V=\big(\frac{\theta}{4}\big)^2\to \infty$ is a limit of extreme
noncommutativity! The new parameter $(1{+}\mathcal{Y})$ corresponds to
a finite wavefunction renormalisation, identified later to decouple
our equations. The parameter $\Lambda^2$ represents an ultraviolet
cut-off which is sent to $\Lambda\to \infty$ in the very end
(continuum limit). In the scaling limit, 
functions of $\frac{|\under{p}|}{\sqrt{V}} =:\mu^2(1+\mathcal{Y})p$
converge to functions of `continuous matrix indices'
$p \in [0,\Lambda^2]$. In the same way, 
$\Gamma_{\under{a}\under{b}}^{ren}$ converges to a function
$\mu^2 \Gamma_{ab}$ with 
$a,b\in [0,\Lambda^2]$, and 
the discrete sum converges to a Riemann integral
\[
\frac{1}{V} \sum_{|\under{p}|=0}^{\mathcal{N}} 
(|\under{p}|+1)  f\big(\tfrac{|\under{p}|}{\sqrt{V}}\big)
\longrightarrow 
\mu^4(1+\mathcal{Y})^2 \int_0^{\Lambda^2} p\,dp\; 
f\big(\mu^2(1+\mathcal{Y})p\big)\;.
\]
This limit makes the restriction to the planar sector (\ref{Gab}) of 
(\ref{Gab-allg}) exact.

After elimination of $\mu^2_{bare}$, but before 
elimination of $Z$, our equation for $\Gamma_{ab}$ becomes 
\begin{align}               
&(Z-1)(1+\mathcal{Y})(a+b) +\Gamma_{ab}
\nonumber
\\
&= -\lambda(1{+}\mathcal{Y})^2
\int_0^{\Lambda^2} \!\!\! p\,dp \;\Big( 
\frac{Z^2}{(a+p)(1{+}\mathcal{Y})+1 -\Gamma_{ap}} 
-\frac{Z^2}{p(1{+}\mathcal{Y})+1 -\Gamma_{0p}} 
\Big)
\nonumber
\\*
& -\lambda(1{+}\mathcal{Y})^2
\int_0^{\Lambda^2}\!\!\! p\,dp \; \Big( 
\frac{Z}{(b+p)(1{+}\mathcal{Y})+1 -\Gamma_{pb}} 
-\frac{Z}{p(1{+}\mathcal{Y})+1 -\Gamma_{p0}} 
  \nonumber
  \\*
  &
\qquad\qquad\qquad 
- \frac{Z}{(b+p)(1{+}\mathcal{Y})+1 -\Gamma_{pb}} 
  \:\frac{\Gamma_{pb}-\Gamma_{ab}}{
(1+\mathcal{Y})(p-a)}
  \nonumber
  \\*
  &
\qquad\qquad\qquad + \frac{Z}{p(1+\mathcal{Y})+1 -\Gamma_{p0}} 
\:  \frac{\Gamma_{p0}}{p(1+\mathcal{Y})}
\Big)\;.
\label{Gamma}
\end{align}
Applying $\frac{d}{db}\big|_{a=b=0}$ we get $Z$ in terms of $\Gamma_{ab}$ 
(and its derivative). Inserted back one gets a highly 
non-linear integro-differential equation. Fortunately we can
reduce the non-linearity by subtracting from (\ref{Gamma}) the same 
equation taken at $b=0$. 
This subtraction eliminates the second line of (\ref{Gamma})
containing $Z^2$. In terms of
$G_{ab}:=\big((a+b)(1{+}\mathcal{Y})+1-\Gamma_{ab}\big)^{-1}$, this
difference equation reads
\begin{align}               
\frac{Z^{-1}}{(1+\mathcal{Y})} \Big(\frac{1}{G_{ab}}-\frac{1}{G_{a0}}\Big)
&= b- \lambda
\int_0^{\Lambda^2} p\,dp \; 
\frac{\frac{G_{pb}}{G_{ab}}-\frac{G_{p0}}{G_{a0}}}{p-a}\;.
\label{GZ}
\end{align}
Differentiation
$\frac{d}{db}\big|_{a=b=0}$ of (\ref{GZ}) yields $Z$ in terms of $G_{ab}$ 
and its derivative. The resulting derivative 
$G'$ can be avoided by adjusting
\[
\mathcal{Y}:=
 -\lambda 
\lim_{b\to 0} 
\int_0^{\Lambda^2} dp \; 
\frac{G_{pb}-G_{p 0}}{b}\;.
\]
This choice leads to
$\displaystyle \frac{Z^{-1}}{(1{+}\mathcal{Y})}=
1-\lambda
\int_0^{\Lambda^2} dp \;G_{p0}$, which is a 
perturbatively divergent integral for $\Lambda\to \infty$. Inserting 
$Z^{-1}$ and $\mathcal{Y}$ back into (\ref{GZ}) we end up in a
\emph{linear} integral equation for the difference function
$D_{ab}:=\frac{a}{b}(G_{ab}-G_{a0})$ to the boundary:
\begin{align}
\Big(\frac{b}{a} + \frac{1}{a G_{a0}} \Big) D_{ab} + 
G_{a0}
= \lambda
\int_0^{\Lambda^2} \!\! dp\; \Big( 
\frac{ D_{pb}
-D_{ab} \frac{G_{p0} }{G_{a0}}}{p-a}
\Big)\;.
\label{D-pre}
\end{align}
The non-linearity restricts to the boundary function $G_{a0}$ where
the second index is put to zero.
Assuming $a \mapsto G_{ab}$ H\"older-continuous, we can pass to
Cauchy principal values. In terms of the \emph{finite Hilbert transform}
\begin{align}
\mathcal{H}^{\Lambda}_a [f(\bullet)]
:= \frac{1}{\pi} 
\lim_{\epsilon\to 0} \Big(\int_0^{a-\epsilon} \!\!\!\! +
\int_{a+\epsilon}^{\Lambda^2}\Big) \frac{f(q)\,dq}{q-a}\;,
\end{align}
the integral equation (\ref{D-pre}) becomes
\begin{align}
\Big(\frac{b}{a} + 
\frac{1 + \lambda \pi a \mathcal{H}_a^{\Lambda}\big[ 
G_{\bullet 0} \big] }{a G_{a 0}}\Big) D_{ab}
-\lambda\pi \mathcal{H}_a^{\Lambda}
\big[ D_{\bullet b} \big]
= -G_{a 0}\;.
\label{D-final}
\end{align}

\subsection{The Carleman solution}

Equation (\ref{D-final}) is a well-known singular integral equation
of Carleman type \cite{Carleman, Tricomi}:
\begin{Theorem}[{\cite{Tricomi}, transformed from $[-1,1]$ to
  $[0,\Lambda^2]$}]
The singular linear integral equation
\[
h(a) y(a) -\lambda\pi 
\mathcal{H}_a^{\Lambda}[y]=f(a)\;,\qquad a\in {]0,\Lambda^2[}\;,
\]
is for $h(a)$ continuous on ${]0,\Lambda^2[}$, H\"older-continuous 
near $0,\Lambda^2$, and $f\in L^p$ for some $p>1$ (determined by 
$\vartheta(0)$ and $\vartheta(\Lambda^2)$)
solved by
\begin{subequations}
\label{Carleman}
\begin{align}
y(a) & 
= \frac{\sin (\vartheta(a))e^{-\mathcal{H}_a^{\Lambda}[\pi-\vartheta]}}{\lambda
  \pi a} \Big(
a \,f(a) e^{\mathcal{H}_a^{\Lambda}[\pi-\vartheta]} \cos(\vartheta(a)) 
\nonumber
\\
& \hspace*{3cm} + 
\mathcal{H}_a^{\Lambda}\Big[e^{\mathcal{H}_\bullet^{\Lambda}[\pi-\vartheta]}
\bullet f(\bullet) \sin(\vartheta(\bullet)) \Big]+ C \Big)
\label{Carleman-a}
\\
&\stackrel{*}{=}
\frac{\sin (\vartheta(a))e^{\mathcal{H}_a^{\Lambda}[\vartheta]}}{\lambda \pi } 
\Big(
f(a) e^{-\mathcal{H}_a^{\Lambda}[\vartheta]} \cos(\vartheta(a)) 
\nonumber
\\
& \hspace*{3cm} + 
\mathcal{H}_a^{\Lambda}\Big[e^{-\mathcal{H}_\bullet^{\Lambda}[\vartheta]}
f(\bullet) \sin(\vartheta(\bullet)) \Big]+
\frac{C'}{\Lambda^2-a}
\Big)\;,
\label{Carleman-b}
\end{align}
\end{subequations}
where $\displaystyle \vartheta(a)=
\di{\raisebox{-1.2ex}{\mbox{\normalsize$\arctan$}}}{
\mbox{\scriptsize$[0,\pi]$}}
\Big(\frac{\lambda\pi}{h(a)}\Big)$, 
$\sin
(\vartheta(a))=\frac{|\lambda\pi|}{\sqrt{(h(a))^2+(\lambda\pi)^2}}
\geq 0$ and  $C,C'$ are arbitrary constants.
\end{Theorem}
The possibility of $C,C'\neq 0$ is due to the fact that the finite
Hilbert transform has a kernel, in contrast to the infinite Hilbert
transform with integration over $\mathbb{R}$.  The two formulae
(\ref{Carleman-a}) and (\ref{Carleman-b}) are formally equivalent, but
the solutions belong to different function classes and normalisation
conditions may (and will) make a choice.

In principle, (\ref{Carleman}) provides the solution $G_{ab}$ of
(\ref{D-final}), where the angle function
\begin{align}
\vartheta_b(a) :=\di{\raisebox{-1.2ex}{\mbox{\normalsize$\arctan$}}}{
\mbox{\scriptsize$[0,\pi]$}}
\mbox{\small$\Bigg(\dfrac{\lambda \pi a}{
b + \frac{1 +  
\lambda \pi a \mathcal{H}^{\Lambda}_a[ G_{\bullet 0} ] }{G_{a 0}}}
\Bigg)$}
\label{vartheta}
\end{align}
plays a key r\^ole. This solution involves multiple Hilbert transforms
which are difficult to control. A better strategy starts from the
observation that the angle (\ref{vartheta}) satisfies, for $b=0$, again a
Carleman type singular integral equation 
\[
\lambda\pi \cot \vartheta_0(a) G_{a0}
-\lambda\pi \mathcal{H}^{\Lambda}_a [G_{\bullet 0} ]= \tfrac{1}{a}
\]
with solution 
\begin{subequations}
\label{Ga0-first}
\begin{align}
G_{a0}
&=\frac{e^{-\mathcal{H}_a^{\Lambda}[\pi-\vartheta_0]}
\sin (\vartheta_0(a))}{\lambda \pi a} \big(
e^{\mathcal{H}_a^{\Lambda}[\pi-\vartheta_0]}\cos(\vartheta_0(a)) 
\nonumber
\\
&\hspace*{4cm}
+
\mathcal{H}_a^{\Lambda}\big[
e^{\mathcal{H}_\bullet^{\Lambda}[\pi-\vartheta_0]}
\sin(\vartheta_0(\bullet))\big]+C
\big)
\\
& \stackrel{*}{=}
\frac{e^{\mathcal{H}_a^{\Lambda}[\vartheta_0]}
\sin (\vartheta_0(a))}{\lambda \pi} \Big(
\frac{e^{-\mathcal{H}_a^{\Lambda}[\vartheta_0]}\cos(\vartheta_0(a))}{a}
\nonumber
\\
&\hspace*{4cm}
+
\mathcal{H}_a^{\Lambda}\Big[
\frac{e^{-\mathcal{H}_\bullet^{\Lambda}[\vartheta_0]}
\sin(\vartheta_0(\bullet))}{\bullet}
\Big]
+\frac{C'}{\Lambda^2-a}
\Big)\;. 
\end{align}
\end{subequations}
Tricomi's identities \cite[\S
  4.4(28+18)]{Tricomi}, which can be arranged as 
\[
e^{\pm \mathcal{H}_a^{\Lambda}[\vartheta_b]}\cos(\vartheta_b(a))
\mp \mathcal{H}_a^{\Lambda}\big[
e^{\pm \mathcal{H}_\bullet^{\Lambda}[\vartheta_b]}\sin(\vartheta_b(\bullet))\big]
=1\;,
\]
and 
rational fraction expansion
$\mathcal{H}_a^{\Lambda}
\big[\frac{f(\bullet)}{\bullet}\big] = \frac{1}{a}\big(
\mathcal{H}_a^{\Lambda} \big[f(\bullet)\big]
- \mathcal{H}_0^{\Lambda}\big[f(\bullet)\big]\big)$ simplify 
(\ref{Ga0-first}) to 
\begin{subequations}
\begin{align}
G_{a0}
&=\frac{e^{-\mathcal{H}_a^{\Lambda}[\pi-\vartheta_0]}
\sin (\vartheta_0(a))}{\lambda \pi a} \big(
C-1\big)
\label{Gab-second-a}
\\
& \stackrel{*}{=}
\frac{e^{\mathcal{H}_a^{\Lambda}[\vartheta_0]}
\sin (\vartheta_0(a))}{\lambda \pi a} \Big(
e^{-\mathcal{H}_0^{\Lambda}[\vartheta_0]}\cos(\vartheta_0(0))
+\frac{C'a}{\Lambda^2-a}
\Big) \;.
\label{Gab-second-b}
\end{align}
\end{subequations}
Both lines are formally equivalent, but we have to guarantee the
normalisation $\lim_{a\to 0} G_{a0}=1$. 
From (\ref{vartheta}) one concludes $\lim_{p\to
  0}\vartheta_0(p)=\left\{ \begin{array}{cl}
    0 & \text{ for } \lambda \geq 0 \\
    \pi & \text{ for } \lambda<0
\end{array}\right\}$. 
Consequently, $e^{-\mathcal{H}_0^{\Lambda}[\vartheta_0]}=
\exp\big(-\frac{1}{\pi} \int_0^{\Lambda^2} \frac{dp}{p}\vartheta_0(p)\big)
\stackrel{\lambda<0}{\longrightarrow 0}$, which means that
(\ref{Gab-second-b}) reduces for $\lambda<0$ to (\ref{Gab-second-a}),
with $C'\mapsto C-1$. Similarly, $\lim_{a\to 0}
e^{-\mathcal{H}_a^{\Lambda}[\pi-\vartheta_0]}
\stackrel{\lambda>0}{=} 0$, so that (\ref{Gab-second-a})
is only consistent with $\lambda<0$.  The normalisation $\lim_{a\to 0}
G_{a0}=1$ leads with $\lim_{a\to 0} \frac{\sin
  \vartheta_0(a)}{|\lambda|\pi a}=1$ to $1-C =
e^{-\mathcal{H}_0^{\Lambda}[\pi-\vartheta_0]}$ in
(\ref{Gab-second-a}), whereas (\ref{Gab-second-b}) stays as it is for
$\lambda>0$. These results can be summarised as follows:
\begin{Lemma}
The angle function 
$\tau_b(a) :=\di{\raisebox{-1.2ex}{\mbox{\normalsize$\arctan$}}}{
\mbox{\scriptsize$[0,\pi]$}}
\Bigg(\dfrac{|\lambda| \pi a}{
b + \frac{1 +
\lambda \pi a \mathcal{H}_a^{\Lambda}[ G_{\bullet 0} ] }{G_{a 0}}}
\Bigg)$ is for $b=0$ reverted to 
\begin{align}
G_{a0} &=
\dfrac{\sin(\tau_0(a))}{|\lambda| \pi a}
\mathrm{e}^{\mathrm{sign}(\lambda)(
\mathcal{H}_0^{\Lambda}[\tau_0(\bullet)]-
\mathcal{H}_a^{\Lambda}[\tau_0(\bullet)])}
\left\{
\begin{array}{@{\!}cl@{}}
1
& \text{ for } \lambda <0 \;,
\\
\big(1{+}\frac{Ca }{\Lambda^2{-}a}\big)
& \text{ for } \lambda > 0 \;,
\end{array}\right.
\label{Ga0}
\end{align}
where $C$ is an arbitrary constant.
\end{Lemma}
Recall that $G_{a0}$ forms the inhomogeneity in the Carleman equation
(\ref{D-final}). We insert (\ref{Ga0}) into the Carleman solution
(\ref{Carleman}) for (\ref{D-final}) and obtain 
with the 
addition theorem
$|\lambda| \pi a\, \sin\big( \tau_d(a)-\tau_b(a)\big)
=  (b-d) \sin \tau_b(a)\sin \tau_d(a)$
after essentially the same steps as in the proof of 
(\ref{Ga0}):
\begin{Theorem}[\cite{Grosse:2014??}]
The full matrix $2$-point function $G_{ab}$ of self-dual $\phi^4_4$-theory on
Moyal space is in the limit $\theta\to \infty$ given in terms of the
boundary $2$-point function $G_{a0}$ by the equation
\begin{align}
G_{ab} =
\dfrac{\sin(\tau_b(a))}{|\lambda|\pi a}
\mathrm{e}^{\mathrm{sign}(\lambda)(
\mathcal{H}_0^{\Lambda}[\tau_0(\bullet)]-
\mathcal{H}_a^{\Lambda}[\tau_b(\bullet)])}
\left\{
\begin{array}{@{\!}cl@{}}
1
& \text{ for } \lambda <0 \;,
\\
\big(1{+}\frac{Ca +bF(b)}{\Lambda^2{-}a}\big)
& \text{ for } \lambda >0\;,
\end{array}\right.
\label{Gab-solution}
\end{align}
where $C$ is an undetermined constant and $b\,F(b)$ an undetermined function
of $b$ vanishing at $b=0$. 
\end{Theorem}
Some remarks:
\begin{itemize}
\item We have proved this theorem in 2012 for $\lambda>0$ under the
  assumption $C'=0$ in (\ref{Carleman-b}), but knew that non-trivial
  solutions of the homogeneous Carleman equation parametrised by
  $C'\neq 0$ are possible. 
That no such term arises for $\lambda<0$
  (if angles are redefined $\vartheta \mapsto \tau$) is a recent
  result \cite{Grosse:2014??}.

\item We expect $C,F$ to be $\Lambda$-dependent
  so that $\big(1{+}\frac{Ca
    +bF(b)}{\Lambda^2{-}a}\big)\stackrel{\Lambda\to
    \infty}{\longrightarrow}  1{+} \tilde{C}a +b\tilde{F}(b)$.

\item An important observation is $G_{ab} \geq 0$, at least for
  $\lambda<0$. This is a truly non-perturbative result; 
  individual Feynman graphs show no positivity at all!

\item As in \cite{Grosse:2009pa}, the equation for $G_{ab}$ can be
  solved perturbatively.  Matching at $\lambda=0$ requires $C,F$ to be
  flat functions of $\lambda$ (all derivatives vanish at zero).
  Because of $\mathcal{H}_a^{\Lambda}[ G_{\bullet 0} ] \stackrel{a\to
    \Lambda^2}{\longrightarrow} -\infty$, the na\"{\i}ve $\arctan$
  series is dangerous for $\lambda>0$.  Unless there are
  cancellations, we expect zero radius of convergence!

\item From (\ref{Gab-solution}) we deduce the 
finite wavefunction renormalisation
\begin{align}
\mathcal{Y}:= -1-\frac{d G_{ab}}{db}\Big|_{a=b=0}
=
\int_0^{\Lambda^2} \!\!\! 
\frac{dp}{(\lambda\pi p)^2 + \big(\frac{1+\lambda\pi p
    \mathcal{H}_p^{\Lambda}[G_{\bullet 0}]}{G_{p0}}\big)^2}
-\left\{\begin{array}{@{\!}c@{\;}l@{}}
0 & \text{for}\;\lambda<0\;,
\\
F(0) & \text{for}\; \lambda>0\;.
\end{array}\right.
\label{Y}
\end{align}

\item The partition function $\mathcal{Z}$ is undefined for $\lambda<0$.  But
the Schwinger-Dyson equations for $G_{ab}$ and for higher functions, 
and with them $\log  \mathcal{Z}$, extend to $\lambda<0$.
These extensions are unique but probably not analytic 
in a neighbourhood of $\lambda=0$.

\end{itemize} 

It remains to identify the boundary function $G_{a0}$. The 
Carleman equation (\ref{D-final}) for $G_{ab}$
 was obtained from the difference (\ref{Gamma})$-$(\ref{Gamma})$_{b=0}$. 
Consequently, (\ref{Gamma})$_{b=0}$ gives the second 
relation between $G_{ab}$ and $G_{a0}$ from which both are
determined. Combining them we obtain a single consistency equation 
for $G_{a0}$, which in terms of 
$\mathcal{T}_a:= |\lambda| \pi a \,\cot \tau_0(a)$ reads \cite{Grosse:2012uv}
\begin{align}               
\mathcal{T}_a
&=1+a+ \lambda\pi a  \mathcal{H}^{\Lambda}_a[1]
\nonumber
\\
&+
\int_0^{\Lambda^2}\!\!  dp \; \bigg( 
\frac{p\, \exp\Big(\mathcal{H}_a^{\Lambda}\Big[
\di{\raisebox{-1.2ex}{\mbox{\small$\arctan$}}}{
\mbox{\scriptsize$[0,\pi]$}}
  \frac{|\lambda| \pi \bullet}{p+\mathcal{T}_\bullet} \Big]\Big)}{
\sqrt{(\lambda\pi a)^2+ (p +\mathcal{T}_a)^2}}
-\frac{p\, \exp\Big(\mathcal{H}_0^{\Lambda}\Big[
\di{\raisebox{-1.2ex}{\mbox{\small$\arctan$}}}{
\mbox{\scriptsize$[0,\pi]$}}
  \frac{|\lambda| \pi \bullet}{p+\mathcal{T}_\bullet} \Big]\Big)}{1+p}
\bigg)\;.
\end{align}
This equation is, unfortunately, of little use. The integrals are
individually divergent for $\Lambda{\to} \infty$ so that we have to
rely on cancellations on which we have no control. 

We compensate this lack by a symmetry argument.  Given the boundary
function $G_{a0}$, the Carleman theory computes the full $2$-point
function $G_{ab}$ via (\ref{Gab-solution}). In particular, we get 
$G_{0b}$ as function of $G_{a0}$. But the $2$-point function is symmetric, 
$G_{ab}=G_{ba}$, and the special case $b=0$ leads to the following 
self-consistency equation:
\begin{Proposition}
The limit $\theta\to \infty$ of $\phi^4_4$-theory on Moyal space is 
determined by the solution of
the fixed point equation $G=TG$,
\begin{align}
G_{b0}
& =  
\frac{\left\{\begin{array}{@{}c@{\;}l@{}}
1 & \text{for}\;\lambda{<}0
\\
1{+}bF(b) & \text{for}\; \lambda{>}0
\end{array}\right\} }{1{+}b}
\exp\Bigg(\!
{-} \lambda \!
\int_0^b \!\!\! dt \! \int_0^{\Lambda^2}  \!\!\!\!\!
\frac{dp}{(\lambda \pi p)^2 
+\big( t {+} \frac{1 {+}\lambda \pi p 
 \mathcal{H}_p^{\Lambda}[G_{\bullet 0}]}{G_{p 0}}\big)^2} 
\Bigg)\;.
\label{master}
\end{align}
\end{Proposition}
At this point we can eventually send $\Lambda \to \infty$.  Any
solution of (\ref{master}) is automatically smooth and (for
$\lambda>0$ but $F=0$) monotonously decreasing.  Any solution of the
true equation (\ref{Gamma}) (without the difference to $b=0$) also
solves the master equation (\ref{master}), but not necessarily
conversely. In case of a unique solution of (\ref{master}) it is
enough to check one candidate.

Existence of a solution of (\ref{master}) is established
(for $\lambda>0$ but $F(b)=0$) by the Schauder fixed point theorem. We
consider the following subset of continuously 
differentiable functions on $\mathbb{R}_+$ vanishing at $\infty$:
\begin{align*}
\mathcal{K}_\lambda:= 
\Big\{ f\in \mathcal{C}^1_0(\mathbb{R}_+)\;  :~  &
f(0)=1\;,\quad 
0 <  f(b)\leq \frac{1}{1+b}\;,\quad
\\*
& 
0\leq -f'(b) \leq \big( \tfrac{1}{1+b}+C_\lambda\big) f(b) \Big\}\;,
\end{align*}
where $C_\lambda$ is defined via $2\lambda P_\lambda^2
(1{+}C_\lambda)e^{C_\lambda P_\lambda}
= 1$ at $P_\lambda = \frac{\exp(-\frac{1}{\lambda\pi^2})}{
\sqrt{1+4\lambda}}$. Then \cite{Grosse:2012uv}:

\begin{enumerate}

\item $\mathcal{K}_\lambda$ is convex,

\item $\overline{T\mathcal{K}_\lambda} \subseteq \mathcal{K}_\lambda$,

\item $(T f)''(b) \leq \big(\frac{23}{4}+\frac{2}{\pi}+\frac{7+8\pi}{2}
\frac{1}{(\lambda\pi^2 P_\lambda)^2}\big) (Tf)(b)$ for any $f\in
\mathcal{K}_\lambda$, 

\item $T: \mathcal{K}_\lambda \to \mathcal{K}_\lambda$ is continuous.

\end{enumerate}
The properties 1.--3.\ imply that $T\mathcal{K}_\lambda$ is relatively
compact in $\mathcal{K}_\lambda$ by a variant of the Arzel\'a-Ascoli
theorem. Together with 4.\ the Schauder fixed point theorem then
guarantees that (\ref{master}) has a solution $G_{a0}\in
\mathcal{K}_\lambda$.

This solution provides $G_{ab}$ via
(\ref{Gab-solution}) and all higher correlation functions via the 
universal algebraic recursion formulae (\ref{GN-real}),  (\ref{GN-B=2-odd}), 
(\ref{GN-B=2-even}), etc, or via the linear equations for the basic 
($N_1{+}\dots{+}N_B)$-point functions such as (\ref{G1+1}) and 
(\ref{G2+2}).
The  recursion formula (\ref{GN-real}) becomes after transition 
to continuous matrix indices 
\begin{align}
G_{b_0\dots b_{N-1}}
&= \frac{(-\lambda)}{(1+\mathcal{Y})^2}
\sum_{l=1}^{\frac{N-2}{2}}
\frac{G_{b_0 b_1 \dots b_{2l-1}} G_{b_{2l}b_{2l+1}\dots b_{N-1}}
- G_{b_{2l} b_1 \dots b_{2l-1}} G_{b_0 b_{2l+1}\dots b_{N-1}}
}{(b_0-b_{2l})(b_1-b_{N-1})}\;.
\label{GN}
\end{align}
It involves the finite wavefunction renormalisation 
$1+\mathcal{Y}=-\frac{d G_{ab}}{db}\big|_{a=b=0}$ given by (\ref{Y}). 
Of particular interest is the 
effective coupling constant
  $\lambda_{\mathit{eff}}=-G_{0000}$. This limit of 
coinciding indices is not so easy; therefore we directly solve 
the integral equation for $G_{a000}$ before using the reality condition. 
We find \cite{Grosse:2012uv}
\begin{align}
\lambda_\mathit{eff}
&= \lambda \bigg\{ 1
+ \frac{\lambda}{(1{+}\mathcal{Y})}
\int_0^{\infty} \!\! dp \;
\frac{
\Big(\dfrac{1-G_{p0}}{(1+\mathcal{Y}) p} - G_{p0}\Big)G_{p0}}{
\big(\lambda\pi pG_{p0}\big)^2 +
\big(1+\lambda\pi p {\mathcal{H}_p}^{\!\!\!\infty}
[G_{\bullet 0}]\big)^2}
\bigg\}\;.
\label{lambdaeff}
\end{align}

The equation for the basic function $G_{ab|cd}$ arising from 
(\ref{G2+2}) is solved in two steps. A first summation over $b \in I$ in 
(\ref{G2+2}) yields after passage to the integral representation a 
Carleman equation for $X_{a|cd}:=\int_0^{\Lambda^2} qdq\; G_{aq|cd}$,
\begin{align*}
&X_{a|cd} \Big\{
1+\lambda \int_0^\infty \!\!\!\!\!dq
\;(G_{aq}-G_{0q})
-\lambda \int_0^\infty  \!\!\!\!\!dq\;
\frac{ G_{aq} \sin \tau_q(a) \cos\big(\tau_q(a)
-\tau_0(a)\big)}{\sin\tau_0(a)}
\Big\}
\\[-1ex]
& + \mathcal{H}_a^{\!\infty}\Big[
\frac{X_{\bullet |cd}}{\pi \bullet}
\int_0^\infty \!\!\!\!\!
q\,dq \;\sin^2 \tau_q(\bullet) G_{aq}\Big]
\nonumber
\\*
&= \lambda \int_0^\infty \!\!\! q\,dq \,
(F_{aq|cdcq}+F_{aq|dcdq})
+ \frac{\lambda}{(1+\mathcal{Y})^2}(G_{acdc} {+} G_{adcd})\;,
\end{align*}
where $F_{ab_1|c_1c_2c_3c_4} := \frac{G_{ab_1c_1c_2c_3c_4}G_{b_1c_3} -
  G_{b_1c_1c_2c_3}G_{ab_1c_3c_4}}{G_{b_1c_1} G_{b_1c_3}}$. Inserted
back into (\ref{G2+2}) gives (after passage to the integral
representation) a familiar Carleman equation for $G_{ab|cd}$ with solution
\begin{align}
G_{ab|cd}
&= F_{ab|cdcb}+F_{ab|dcdb}
\nonumber
\\*
& - \frac{\sin\tau_b(a)}{\lambda\pi a} \cos \tau_b(a)
G_{ab} X_{a|cd}
-G_{ab}
\mathcal{H}_a^{\!\infty}\Big[ \frac{\sin^2 \tau_b(\bullet)}{
\lambda\pi \bullet}   X_{\bullet| cd}
\Big]\;.
\end{align}
The $(2{+}2)$-point function $G_{ab|cd}$ turns out to be the most interesting 
part of the 4-point function in position space (see sec.~\ref{sec:Schwinger}).

\subsection{Perturbation theory}

The master equation (\ref{master}) can, for $F(b)\equiv 0$, be
iteratively solved. To lowest order one has
$G_{a0}=\frac{1}{1+a} +\mathcal{O}(\lambda)$, from which the next
order becomes
\begin{align}
G_{a0} = \frac{1}{1+a}-\lambda \frac{\log(1+a)}{(1+a)}
+ \mathcal{O}(\lambda^2)\;.
\label{Ga0-pert}
\end{align}
If we put in $G_{a0}=\frac{1}{(1+a)^{1+\lambda}}
{+}\mathcal{O}(\lambda^2)$ the index $a\mapsto \frac{p^2}{\mu^2}$, see
(\ref{Schwinger-final}), we get
\begin{align*}
\int_{\mathbb{R}^4}
\frac{dp}{(2\pi\mu)^4} 
\;e^{\mathrm{i}p(x-y)}G_{\frac{p^2}{\mu^2}0}
&=\frac{2^ {-\lambda}}{4\pi^2\Gamma(1+\lambda)} 
\frac{K_{1-\lambda} (\mu\|x-y\|)}{
(\mu\|x-y\|)^{1-\lambda}} 
\\
&\stackrel{x-y\to 0}{\longrightarrow}
\frac{2^{-2\lambda} \Gamma(1-\lambda)}{4\pi^2\Gamma(1+\lambda)} 
\frac{1}{(\mu\|x-y\|)^{2-2\lambda}}\;,
\end{align*}
where $K_\nu(x)$ is the modified Bessel function.
We thus conclude that the anomalous dimension is $\eta=-2\lambda$,
i.e.\ negative for the stable sign $\lambda>0$ of the coupling
constant. We shall see in the next section that this result excludes a
Wightman theory for $\lambda>0$.  It is worthwhile to mention that this
wrong sign is a consequence of renormalisation. The divergent bare
$2$-point function would lead to the opposite sign. Removing the
divergence at $a=0$ overcompensates for $a>0$ and gives
$\eta=-2\lambda$. In two dimensions, $\eta$ would be non-negative for
$\lambda>0$.

From (\ref{Ga0-pert}) we get:
\begin{itemize}
\item Hilbert transform: 
$\lambda\pi \mathcal{H}_a^{\!\infty}[G_{\bullet 0}]
= -\lambda \frac{\log(a)}{1+a} +
\mathcal{O}(\lambda^2)$,

\item angle function: $\tau_b(a) = \frac{|\lambda|\pi a}{1+a+b}
\Big(1- \lambda \frac{
(1+a)\log(1+a)-a\log a}{(1+a+b)}\Big)+\mathcal{O}(\lambda^3)$,

\item wavefunction renormalisation: $1+\mathcal{Y}
=-\frac{d G_{a0}}{da}\big|_{a=0}
=1+\lambda+\mathcal{O}(\lambda^2)$.

\end{itemize}
Inserted into (\ref{Gab-solution}) one finds
\begin{align}
G_{ab} = \frac{1}{1+a+b}
-\lambda
\frac{(1+a)\log(1+a)
+(1+b)\log(1+b) }{(1+a+b)^2}+\mathcal{O}(\lambda^2)\;.
\label{Gab-pert}
\end{align}

This result coincides with renormalised 1-loop ribbon graph
computation. From the action functional (\ref{Sphi}) one obtains in
the infinite volume limit to continuous matrix indices the following
Feynman rules:
\begin{itemize}
\item 
\parbox{2cm}{\begin{picture}(20,5)
\put(6,0){\includegraphics[width=8mm]{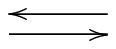}}
\put(10,-2){\mbox{\scriptsize$b$}}
\put(10,3){\mbox{\scriptsize$a$}}
\end{picture}}\quad
$=\dfrac{1}{1+(a+b)(1+\mathcal{Y})}$

\item 
\parbox{2cm}{\begin{picture}(18,14)
\put(0,0){\includegraphics[width=1.5cm]{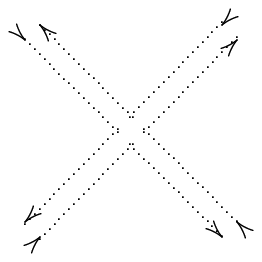}}
\end{picture}}\quad
$=-Z^2\lambda$\quad(index conserved at every corner)

\item \parbox{2cm}{\begin{picture}(18,12)
\put(7,7){\circle{12}}
\put(6,6){\mbox{\scriptsize$p$}}
\end{picture}}\quad
$\displaystyle = (1{+}\mathcal{Y})^2 \int_0^{\Lambda^2} p\,dp$\quad
for every closed face

\end{itemize}

To lowest order we have
$G_{ab} =\dfrac{1}{1+(a+b)(1+\mathcal{Y})-\Gamma_{ab}^{ren}}$, where 
$\Gamma^{rn}_{ab}$ is the Taylor remainder of
\begin{subequations}
\begin{align}
\Gamma_{ab} &= 
\parbox{1.6cm}{\begin{picture}(15,21)
\put(0,0){\includegraphics[width=1.5cm]{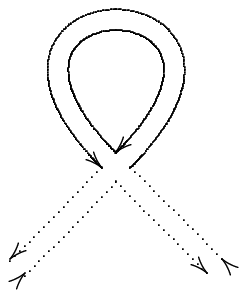}}
\put(7.5,3){\mbox{\scriptsize$b$}}
\put(7.5,13){\mbox{\scriptsize$p$}}
\put(3,8){\mbox{\scriptsize$a$}}
\put(11,8){\mbox{\scriptsize$a$}}
\end{picture}} + 
\parbox{1.6cm}{\begin{picture}(15,21)
\put(0,0){\includegraphics[width=1.5cm]{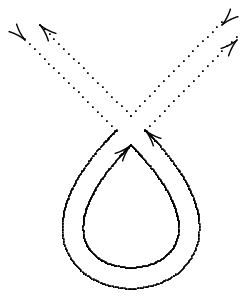}}
\put(7.5,5){\mbox{\scriptsize$p$}}
\put(7.5,15){\mbox{\scriptsize$a$}}
\put(3,11){\mbox{\scriptsize$b$}}
\put(11,11){\mbox{\scriptsize$b$}}
\end{picture}} +\mathcal{O}(\lambda^2)
\nonumber
\\
&= \int_0^{\Lambda^2} \!\!\! p\,dp \;\frac{(-\lambda)}{1+a+p} + 
\int_0^{\Lambda^2} \!\!\!p\,dp\; \frac{(-\lambda)}{1+b+p} + 
\mathcal{O}(\lambda^2)\;,
\nonumber
\\
&= \Gamma^{ren} + \bigg(
\int_0^{\Lambda^2} \!\!\! p\,dp \; \frac{(-\lambda)}{1+p} 
+ \underbrace{\int_0^{\Lambda^2} \!\!\! p\,dp\; \frac{(+\lambda)a}{
(1+p)^2}}_{(Z-1)a}
+ \big(a\mapsto b\big)  +\mathcal{O}(\lambda^2)\bigg)\;,
\\[-0.3ex]
\Gamma^{ren}_{ab} & 
=(-\lambda) \int_0^{\Lambda^2} \!\!\! p\,dp \,
\Big(\frac{1}{1{+}a{+}p}-\frac{1}{1{+}p} +\frac{a}{(1{+}p)^2}\Big) 
+\big(a\mapsto b\big) +\mathcal{O}(\lambda^2)\;,
\end{align}
\end{subequations}
in agreement with (\ref{Gab-pert}). There is no doubt that the fixed point
solution for $G_{a0}$ and the Carleman solution for $G_{ab}$ capture
the resummation of infinitely many renormalised Feynman graphs!

From (\ref{GN}) and $\mathcal{Y}=\lambda
+\mathcal{O}(\lambda^2)$ we obtain for the $4$-point function
\begin{align}
G_{abcd} & =\frac{(-\lambda)}{(1+\mathcal{Y})^2}
\frac{G_{ab}G_{cd}-G_{ad}G_{cd}}{(a-c)(b-d)} =: 
G_{ab}G_{bc}G_{cd}G_{da} (-\Gamma_{abcd})\;,
\nonumber
\\
\Gamma_{abcd} & =\lambda\Big(1
-\lambda
\frac{a- (1+a)\log(1+a)-c+(1+c)\log(1+c) }{a-c}
\nonumber
\\
&\quad -\lambda
\frac{b- (1+b)\log(1+b)-d+(1+d)\log(1+d) }{b-d}
\Big) + \mathcal{O}(\lambda^3)\;,
\end{align}
which agrees with 
\begin{align}
\Gamma_{abcd} & = - \parbox{1.8cm}{\begin{picture}(18,18)
\put(0,0){\includegraphics[width=1.5cm]{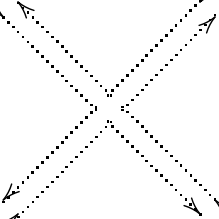}}
\put(7,3.5){\mbox{\footnotesize{$a$}}}
\put(10.5,7){\mbox{\footnotesize{$b$}}}
\put(7,10.5){\mbox{\footnotesize{$c$}}}
\put(3.5,7){\mbox{\footnotesize{$d$}}}
\end{picture}}
 - \parbox{2.8cm}{\begin{picture}(28,18)
\put(0,3){\includegraphics[width=2.5cm]{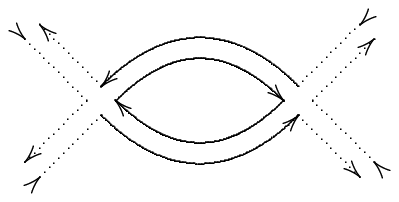}}
\put(5,4){\mbox{\footnotesize{$a$}}}
\put(18,4){\mbox{\footnotesize{$a$}}}
\put(22.5,8){\mbox{\footnotesize{$b$}}}
\put(6,12.5){\mbox{\footnotesize{$c$}}}
\put(18,12.5){\mbox{\footnotesize{$c$}}}
\put(0.5,8){\mbox{\footnotesize{$d$}}}
\put(12,8){\mbox{\footnotesize{$p$}}}
\end{picture}}
 - \parbox{1.1cm}{\begin{picture}(10,18)
\put(0,-4){\rotatebox{90}{\includegraphics[width=2.5cm]{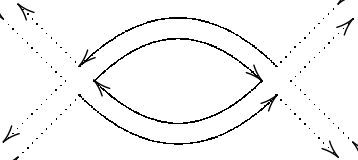}}}
\put(5,-2){\mbox{\footnotesize{$a$}}}
\put(8.5,1){\mbox{\footnotesize{$b$}}}
\put(8.5,15){\mbox{\footnotesize{$b$}}}
\put(5,18){\mbox{\footnotesize{$c$}}}
\put(1,1){\mbox{\footnotesize{$d$}}}
\put(1,15){\mbox{\footnotesize{$d$}}}
\put(5,8){\mbox{\footnotesize{$p$}}}
\end{picture}} + \mathcal{O}(\lambda^3)
\nonumber
\\
&= -(-\lambda)\underbrace{\Big(1+2\lambda 
\int_0^{\Lambda^2} \!\!\! \frac{p\,dp}{(1+p)^2} \Big)}_{= Z^2 +\mathcal{O}(\lambda^2)}
\nonumber
\\
& 
-(-\lambda)^2 \!\!\int_0^{\Lambda^2} \!\!\! 
\frac{ p\,dp}{(1{+}p{+}a)(1{+}p{+}c)}
-(-\lambda)^2 \!\!\int_0^{\Lambda^2} \!\!\! 
\frac{ p\,dp}{(1{+}p{+}b)(1{+}p{+}d)}\;.
\end{align}
The singularities of $Z^2$ and of 
the one-loop 4-point graphs cancel exactly!

\subsection{Computer simulations \cite{Grosse:2014??}}
\label{sec:computer}

A numerical investigation of (\ref{master}), for $F(b)\equiv 0$, 
reveals interesting properties of the $\phi^4_4$-theory on Moyal space.
We approximate $G_{a0}$ as piecewise linear function
on $[0,\Lambda^2]$ sampled according to a geometric progression and
view (\ref{master}) as iteration $G^{n+1}_{a0}=(T G^n)_{a0}$ for some
initial function $G^0$. In this way we find numerically that $T$
satisfies, for any $\lambda \in \mathbb{R}$, the assumptions of the
Banach fixed point theorem for Lipschitz functions on $[0,\Lambda^2]$,
i.e.\ $T$ is contractive and $(G^{n})$ converges to a fixed point
which approximates $G_{a0}$. Whereas $(G^{n})$
converges for any sign of $\lambda$ (without discontinuity at
$\lambda=0$), the necessary consistency condition $G_{ab}=G_{ba}$ for
(\ref{Gab-solution}) turns out to be maximally violated for
$\lambda>0$ (assuming $C=0=F(b)$) and satisfied (within numerical
error bounds) for $\lambda\leq 0$. The observed relative asymmetry
$\sup_{a,b} \big|\frac{G_{ab}-G_{ba}}{G_{ab}+G{ba}}\big|$ of nearly
$100\,\%$ for $\lambda>0$ signals that the parameters $C,F(b)$ in
(\ref{Gab-solution}) which reflect the non-trivial solution of the
homogeneous Carleman equation are definitely non-zero. Taking
$C,F(b)\neq 0$ for $\lambda>0$ into account is not feasible at the
moment so that our numerical results are reliable only for
$\lambda\leq 0$. For $\lambda=10^7$ and only $2000$ sample points in
$[0,\Lambda^2]$, the relative asymmetry for $\lambda\leq 0$ is of the
order of $5\,\%$.

\begin{figure}[h]
\begin{picture}(150,117)
\put(60,2){\colorbox{palered}{\rule{0mm}{115mm}\hspace*{5.65cm}\rule{0mm}{115mm}}}
\put(37.2,2){\colorbox{palegreen}{\rule{0mm}{115mm}\hspace*{2.1cm}\rule{0mm}{115mm}}}
\put(1,2){\colorbox{palered}{\rule{0mm}{115mm}\hspace*{3.4cm}\rule{0mm}{115mm}}}
  \put(0,0){\includegraphics[width=12cm,
       viewport=0 0 288 288]{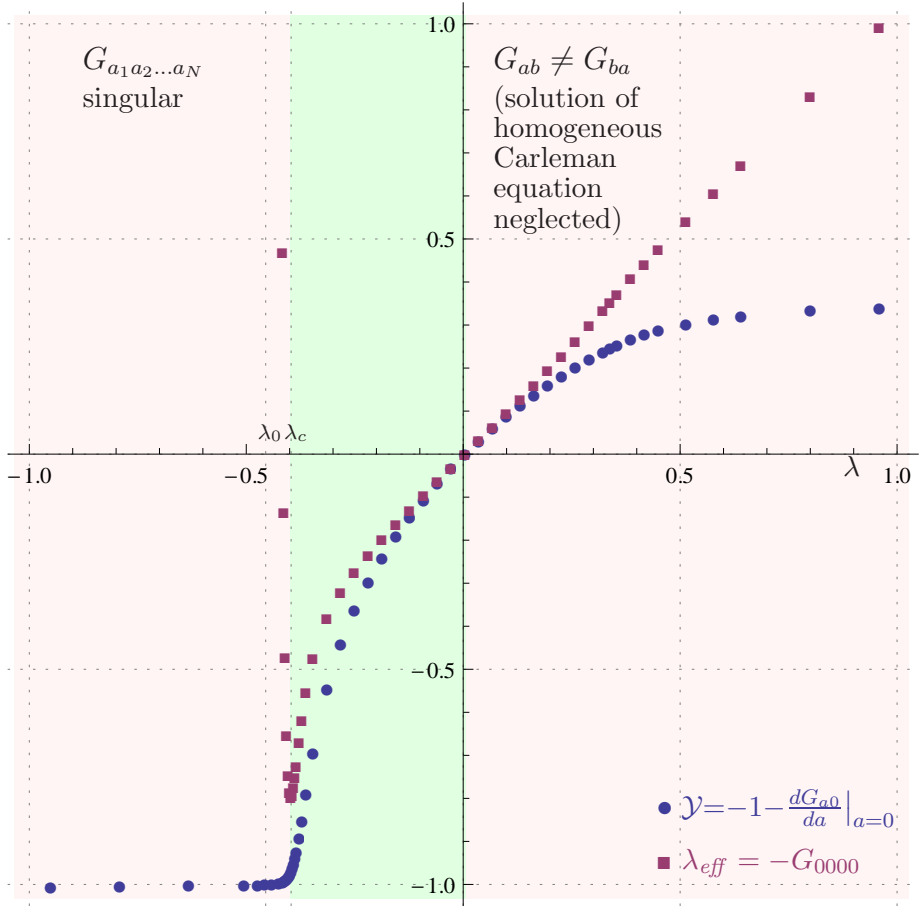}}
 \put(110,57){\mbox{\small$\lambda$}}
 \put(64,111){\mbox{{$G_{ab}\neq G_{ba}$}}}
 \put(64,106){\mbox{\small{(solution of}}}
 \put(64,102){\mbox{\small{homogeneous}}}
 \put(64,98){\mbox{\small{Carleman}}}
 \put(64,94){\mbox{\small{equation}}}
 \put(64,90){\mbox{\small{neglected)}}}
 \put(10,111){\mbox{{$G_{a_1a_2\dots a_N}$}}}
 \put(10,106){\mbox{\small{singular}}}
  \put(33,62){\mbox{\scriptsize$\lambda_0$}}
  \put(36.5,62){\mbox{\scriptsize$\lambda_c$}}
  \put(85.5,12){\mbox{\textcolor{mathblue}{
  \textbullet{\small$\;\mathcal{Y}{=} {-}1{-}\frac{dG_{a0}}{da}\big|_{a=0}$}}}}
  \put(85.5,5){\mbox{\textcolor{mathred}{{\tiny$\blacksquare$}
             {\small$\lambda_{\mathit{eff}}=-G_{0000}$}}}}
\end{picture}
\caption{$\mathcal{Y},\lambda_{\mathit{eff}}$ based on $G_{a0}$ for 
$\Lambda^2{=}10^7$ with 2000 sample points.\label{fig1}}
\end{figure}

\begin{figure}[h]
\begin{picture}(150,42)
  \put(0,0){\includegraphics[width=7cm,
            viewport=0 0 288 175]{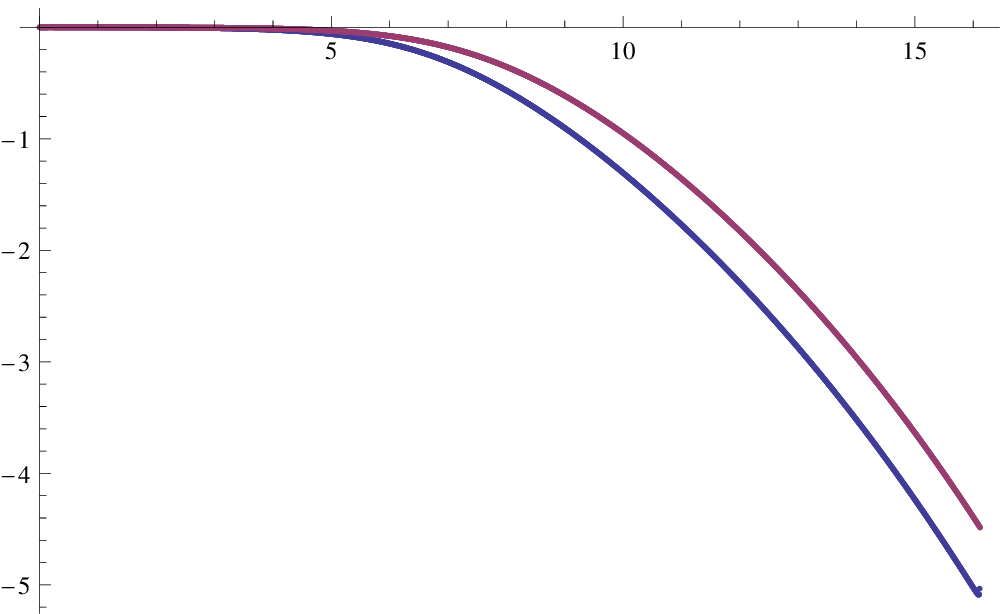}}
  \put(50,39){\mbox{\tiny$\log(1{+}a)$}}
  \put(51,29){\mbox{\textcolor{mathred}{\footnotesize$\log G_{a0}$}}}
  \put(45,15){\mbox{\textcolor{mathblue}{\footnotesize$\log G_{aa}$}}}
  \put(10,7){\fbox{\mbox{\small$\lambda=-0.477$}}}
  \put(75,0){\includegraphics[width=7cm,
              viewport=0 0 288 175]{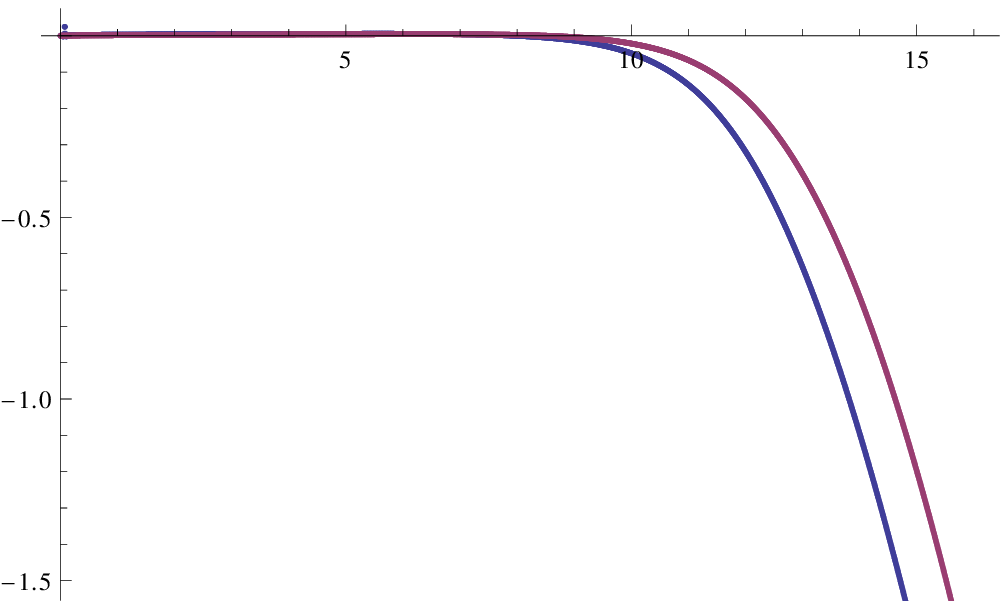}}
  \put(127,38){\mbox{\tiny$\log(1{+}a)$}}
  \put(133,29){\mbox{\textcolor{mathred}{\footnotesize$\log G_{a0}$}}}
  \put(121,17){\mbox{\textcolor{mathblue}{\footnotesize$\log G_{aa}$}}}
  \put(86,7){\fbox{\mbox{\small$\lambda=-0.796$}}}
\end{picture}
\caption{Plots of $\log G_{a0}$ and $\log G_{aa}$ over $\log (1+a)$
  for $\lambda<\lambda_c$. \label{fig2}}
\end{figure}

The most striking outcome of our computer simulations concerns the
finite wavefunction renormalisation $(1+\mathcal{Y})$ given by (\ref{Y}). 
Figure~\ref{fig1} shows both $\mathcal{Y}$
and the effective coupling constant $\lambda_{\mathit{eff}}$ given by
(\ref{lambdaeff}) as functions of $\lambda$.
We find clear evidence for a second-order phase transition: 
$\mathcal{Y}'$ is discontinuous at $\lambda_c=-0.396$, and we have
in reasonable approximation a critical behaviour 
\begin{align}
1+\mathcal{Y}= \left\{
\begin{array}{cl} A (\lambda-\lambda_c)^\alpha & \text{ for }
    \lambda\geq \lambda_c\;,
\\
0 &  \text{ for }
    \lambda < \lambda_c\;,
  \end{array}\right.
\label{critic}
\end{align}
for some $A,\alpha>0$. To be precise, we find $1+\mathcal{Y}=0$ only
at $\lambda_0{=}-0.455$, but this seems to be due to the
discretisation. Of course, there cannot be a discontinuity in
$\mathcal{Y}'$ for finite $\Lambda$, but Figure~\ref{fig1} is strong
support for a critical behaviour (\ref{critic}) in the limit
$\Lambda^2\to \infty$. It is worthwhile to mention that nothing
particular happens at the expected pole
$\lambda_b=-\frac{1}{72}=0.014$ of Borel resummation! Since
$1+\mathcal{Y}=0$ (within numerical error bounds) in the phase
$\lambda<\lambda_c$, we see from (\ref{GN}) that higher $N$-point
functions will not exist for $\lambda < \lambda_c$.  Most
surprisingly, as we discuss at the end of section~\ref{sec:reflect}, a
key property of the Schwinger 2-point function $S_c(x,y)$ in position
space is precisely realised in $[\lambda_c,0]$, not outside!  In fact,
as shown in Figure~\ref{fig2},
one has in reasonable approximation $G_{ab}=0$ for $0\leq a,b\leq
\Lambda_0^2$, where $\Lambda_0^2$ increases with
$\lambda_c-\lambda>0$. This could leave the possibility of meaningful
higher functions (\ref{GN}) for matrix indices $0\leq a_i \leq
\Lambda_0^2$, but not for larger indices. Such a picture could have
the interpretation of a maximal momentum cut-off of the Euclidean
particles in the phase $\lambda<\lambda_c$.

\section{Schwinger functions and reflection positivity}
\label{sec:Schwinger}

In the previous section we have constructed the connected matrix correlation
functions $G_{|\under{q}^1_1\dots\under{q}^1_{N_1}|\dots
|\under{q}^B_1\dots\under{q}^B_{N_B}|}$ of the ($\theta{\to}
\infty$)-limit of $\phi^4_4$-theory on Moyal space. These functions
arise from the topological expansion (\ref{logZ}) of the free energy 
\begin{align}
\log\frac{\mathcal{Z}[J]}{\mathcal{Z}[0]}= \! \sum_{B=1}^\infty
\sum_{1\leq N_1 \leq \dots \leq N_B}^\infty \!\!\!\!\!
\frac{(V\mu^4)^{2-B}}{S_{N_1\dots N_B}} 
\!\! \sum_{\under{q}^{\beta}_i \in \mathbb{N}^2} \!\!\!
G_{|\under{q}^1_1\dots\under{q}^1_{N_1}|\dots
|\under{q}^B_1\dots\under{q}^B_{N_B}|} 
\prod_{\beta=1}^B \!\frac{1}{N_\beta}\Big(
\frac{J_{\under{q}^\beta_1\under{q}^\beta_2}}{\mu^3}{\cdots} 
\frac{J_{\under{q}^\beta_{N_\beta}\under{q}^\beta_1}}{\mu^3}\Big) .
\label{logZ-Moyal}
\end{align}
Since $\lim_{V\mu^4 \to \infty} G_{|\under{q}^1_1\dots\under{q}^1_{N_1}|\dots
|\under{q}^B_1\dots\under{q}^B_{N_B}|}$ is finite,
the limit 
$\lim_{V\to \infty}
\frac{1}{V\mu^4} \log\frac{\mathcal{Z}[J]}{\mathcal{Z}[0]}$ 
of the naturally expected  free energy density 
removes (in addition to the removal of higher-genus contributions) 
all contributions from $B\geq 2$. As shown in previous sections, 
this planar limit is an exactly solvable and (without any doubt) 
non-trivial matrix model.

\subsection{Schwinger functions}

We are interested here in another limit to Schwinger functions
\cite{Schwinger:1959zz} in position space. For this end we revert the
matrix representation (\ref{matrixbasis}) and take the infinite volume
limit $V\mu^4\to \infty$, where we carefully have to pass to
densities.  Absolute position $x\in \mathbb{R}^4$ have no meaning,
only $\mu x$ can be used. This means that we consider
(up to a factor discussed below)
\begin{align*}
&\big\langle
  \phi(\mu x_1)\dots 
\phi(\mu x_N)\big\rangle
\equiv \!\!\!\!
\sum_{\under{m}_1,\under{n}_1,\dots,
\under{m}_N,\under{n}_N \in \mathbb{N}^2}
\!\!\!\!\!\!\!\!\!\!\!\! f_{\under{m}_1\under{m}_2}(\mu x_1) \cdots
f_{\under{m}_N\under{m}_N}(\mu x_N) 
\big\langle \Phi_{\under{m}_1\under{n}_1} \dots
\Phi_{\under{m}_N\under{n}_N}\big\rangle ,
\end{align*}
where the matrix correlation functions $\big\langle
\Phi_{\under{m}_1\under{n}_1} \dots
\Phi_{\under{m}_N\under{n}_N}\big\rangle$ are obtained by derivatives
of (\ref{logZ-Moyal}) with respect to
$J_{\under{m}_1\under{n}_1},\dots , J_{\under{m}_N\under{n}_N}$. We
shall see in this section that the additional index summation over $
\under{m}_i,\under{n}_i \in \mathbb{N}^2$ gives a meaningful limit
only if we redefine the volume factor in the free energy density to
$\mathcal{F}=\frac{1}{(V\mu^4)^2}\log\frac{\mathcal{Z}[J]}{\mathcal{Z}[0]}$.
The occurrence of $V^2$ as the volume has its origin in the spectral
geometry of the Moyal plane with harmonic propagation
\cite{Gayral:2011vu, Grosse:2007jy} which has a finite volume
$(\frac{V}{\Omega})^2$.
\begin{Definition}
The connected Schwinger functions associated with the action
(\ref{GW}) are 
\begin{align}
&\mu^N S_c(\mu x_1,\dots ,\mu x_N) 
\nonumber
\\
&:=\!\!\! \lim_{V\mu^4\to \infty} \!\!\!\!\!\!\!
\sum_{\under{m}_1,\under{n}_1,\dots,
\under{m}_N,\under{n}_N \in \mathbb{N}^2}
\!\!\!\!\!\!\!\!\!\!\!\! 
f_{\under{m}_1\under{n}_1}(\mu x_1)
\cdots f_{\under{m}_N\under{n}_N}(\mu x_N)
\frac{\mu^{4N} \partial^N \mathcal{F}[J]}{\partial J_{\under{m}_1\under{n}_1}
{\dots} \partial J_{\under{m}_N\under{n}_N}}\bigg|_{J=0}\;,
\\
&\mathcal{F}[J]:=\frac{1}{64\pi^2 V^2\mu^8}\log\left(
\dfrac{\int \mathcal{D}[\Phi] \;
e^{-S[\Phi]+ V \sum_{\under{a},\under{b}\in \mathbb{N}^2} \Phi_{\under{a}\under{
b}}
J_{\under{b}\under{a}}} }{
\int \mathcal{D}[\Phi] \; e^{- S[\Phi]}}
\right)_{\di{Z\mu_{bare}^2\mapsto \mu^2}{Z\mapsto (1+\mathcal{Y})}}\;,
\nonumber
\end{align}
where $S[\Phi]$ is given by (\ref{Sphi}) and
$f_{\under{m}\under{n}}$ by (\ref{matrixbasis}).
By $(\quad)_{\di{Z\mu_{bare}^2\mapsto \mu^2}{Z\mapsto
    (1+\mathcal{Y})}}$ we symbolise the renormalisation 
of sec.~\ref{sec:renorm}. 
\end{Definition}
Note that by construction the $J$-derivatives, and hence the Schwinger
functions, are fully symmetric in $\mu x_1,\dots,\mu x_N$. Applying the 
$J$-derivatives the the topological expansion (\ref{logZ-Moyal}) into
$J$-cycles produces an $f_{\under{m}\under{n}}$-cycle for each of the
$B$ boundary components: 
\begin{align*}
S_c(\mu x_1,{\dots},\mu x_N)
&= \lim_{V\mu^4\to \infty}
\frac{1}{64\pi^2}  \sum_{N_1+\dots+N_B=N}
\sum_{\under{q}^{\beta}_i \in \mathbb{N}^2} 
G_{|\under{q}^1_1\dots\under{q}^1_{N_1}|\dots
|\under{q}^B_1\dots\under{q}^B_{N_B}|}
\\*
& \times\!\!\!
\sum_{\sigma \in \mathcal{S}_N} \prod_{\beta=1}^B 
\frac{
f_{\under{q}_1\under{q}_2}(
\mu x_{\sigma(N_1{+}{\dots}{+}N_{\beta{-}1}{+}1)})
{\cdots}
f_{\under{q}_{N_\beta}\under{q}_1}(\mu
x_{\sigma(N_1{+}{\dots}{+}N_\beta)}) 
}{V\mu^4 N_\beta} \;.
\end{align*}

We compute the sum over the indices $\under{q}^{\beta}_i \in
\mathbb{N}^2$ by Laplace-Fourier transform of $G$.  For that we
temporarily assume that $G$ has, for every boundary component, a
representation as Laplace transform in the total sum of index norms
and Fourier transform in differences of index norms.  This transform
will be reverted in the end so that the analyticity assumption is not
necessary (future analytic continuation to Minkowski space would imply
representation as Laplace transform):
\begin{align}
&G_{|\under{q}^1_1 \dots \under{q}^1_{N_1}|\dots|
\under{q}^B_1 \dots \under{q}^B_{N_B}|}
\nonumber
\\
&=\int_{\mathbb{R}_+^B} \!\!\! d(t^1,\dots, t^B)
\int_{\mathbb{R}^{N-B}} \!\!\!
d(\omega^1_1,\dots,\omega^1_{N_1-1},\dots ,
\omega^B_1,\dots,\omega^B_{N_B-1}) 
\nonumber
\\
& 
\qquad \times 
\mathcal{G}(t^1,\omega^1_1,\dots,\omega^1_{N_1{-}1}|
\dots | t^B,\omega^B_1,\dots,\omega^B_{N_B{-}1})
\nonumber
\\
&\qquad \times \prod_{\beta=1}^B 
\exp\bigg( {-}\frac{t^\beta}{\sqrt{V\mu^4}}\sum_{i=1}^{N_\beta} 
|\under{q}^\beta_i|
+\frac{\mathrm{i}}{\sqrt{V\mu^4}} 
\sum_{i=1}^{N_\beta-1} \omega^\beta_i
(|\under{q}^\beta_i|-|\under{q}^\beta_{i+1}|)\bigg)\;.
\label{calG}
\end{align}
Note that the 1-norms 
$|\under{q}^\beta_i|= q^\beta_{i,1}{+}
q^\beta_{i,2}$ imply a factorisation of the exponential, 
$\exp(\dots){=}\mbox{\small$\displaystyle\prod_i$} 
\big(z^\beta_i(t^\beta,\vec{\omega}^\beta)
\big)^{\!q^\beta_{i,1}}
\big(z^\beta_i(t^\beta,\vec{\omega}^\beta)
\big)^{\!q^\beta_{i,2}}$.

For every boundary component $\beta=1,\dots,B$, we thus need to compute
\begin{align}
&\sum_{q_1,\dots,q_{N'}=0}^\infty 
\frac{
f_{q_1q_2}(\mu \vec{y}_1) \cdots 
f_{q_{N'}q_1}(\mu \vec{y}_{N'}) 
}{\sqrt{V\mu^4 N'}} z_1^{q_1}\cdots
z_{N'}^{q_{N'}}
\nonumber
\\*
&= 2^{N'} \!\!\!\!\!\!\!\!
\sum_{q_1,\dots,q_{N'}=0}^\infty \!\!\!\!\!\!
e^{-\frac{1}{2}(r_1+\dots+r_{N'})}
\frac{
L_{{q_1}}^{{q_2}-{q_1}}(r_1) 
\cdots 
L_{{q_{N'}}}^{{q_1}-{q_{N'}}}(r_{N'})}{ 
\sqrt{V\mu^4 N'}} 
({-}\tilde{z}_1)^{{q_1}} \cdots ({-}\tilde{z})_{N'}^{{q_{N'}}}\;,
\label{sum-f}
\end{align}
where $r_i=\frac{\mu^2|\vec{y}_i|^2}{\sqrt{4V\mu^4}}$ and 
$\tilde{z}_j =
\frac{\vec{y}_{j-1}}{\vec{y}_j}
\exp(-\frac{t-\mathrm{i}(\omega_j-\omega_{j-1}) }{\sqrt{V\mu^4}})$,
with 
$\vec{y}_i\in \mathbb{C}$, $\vec{y}_0\equiv\vec{y}_{N'}$ and 
$\omega_0=\omega_{N'}\equiv 0$. One has
\begin{Lemma}[\cite{Grosse:2013iva}]
For $|\tilde{z}_j|<1$, a cyclic product of Laguerre polynomials (i.e.\
$N'+j\equiv j$) is summed to 
\begin{align}
\sum_{q_1,\dots,q_{N'}=0}^\infty
\prod_{j=1}^{N'} (-\tilde{z}_j)^{q_j}\, L_{q_j}^{q_{j+1}-q_j}(r_j)
=\frac{\displaystyle 
\exp\Bigg(\!\!{-}\frac{
\mbox{\small$\sum_{j,k=1}^{N'} {r_j}
({-}\tilde{z}_{k+j})\cdots ({-}\tilde{z}_{N'+j})$}
}{{1-(-\tilde{z}_1)\cdots (-\tilde{z}_{N'})}}
\Bigg)}{{1-(-\tilde{z}_1)\cdots (-\tilde{z}_{N'})}}\;.
\label{Lemma-Laguerre}
\end{align}
\end{Lemma}
The denominators in (\ref{Lemma-Laguerre}) become
\[
{1{-}({-}\tilde{z}_1)\cdots ({-}\tilde{z}_{N'})
{=}1{-}({-}1)^{N'}\exp\Big({-}\frac{N't}{\sqrt{V\mu^4}}\Big)}
\stackrel{{V\mu^4\to \infty}}{\longrightarrow} 
\left\{ \!\!\!\! 
\begin{array}{cl} \frac{N't}{\sqrt{V\mu^4}}  & \text{for $N'$
      even}\;, \\
2 & \text{for $N'$ odd}\;.
\end{array}\right. 
\]
Together with the prefactor $\frac{1}{\sqrt{V\mu^4N'}}$, the sum
(\ref{sum-f}) converges for $V\mu^4\to \infty$ to zero if $N'$ is odd,
whereas if $N'$ is even the limit is non-zero and finite, depending
only on $t$ but no longer on $\omega_j$. Recombining the two
$\mathbb{N}^2$-components we produce factors 
$\displaystyle \frac{\exp\big({-}\frac{\|\mu X\|^2}{2N'{t}}\big)}{
  (N't)^2} = \int_{\mathbb{R}^4} \frac{dp}{4\pi^2 \mu^4}
e^{-\mathrm{i}\langle \frac{p}{\mu},\mu X\rangle} \exp\Big( -\frac{N'
  t \|p\|^2}{2\mu^2} \Big) $ for every even $N'$. Altogether we arrive at
\begin{align*}
&\lim_{V\mu^4\to \infty} \sum_{\under{q}_1,\dots,\under{q}_{N'}
  \in \mathbb{N}^2} 
\frac{
f_{\under{q}_1\under{q}_2}(\mu x_1) \cdots 
f_{\under{q}_{N'}\under{q}_1}(\mu x_{N'}) 
}{V\mu^4 N'} z_1^{q_{1,1}+q_{1,2}}\cdots
z_{N'}^{q_{N',1}+q_{N',2}}
\\
&=\left\{
\begin{array}{cl} \!\!
\displaystyle
\frac{4^{N'}}{N'} 
 \int_{\mathbb{R}^4} \frac{dp}{4\pi^2 \mu^4} e^{-\mathrm{i}\langle
  \frac{p}{\mu},\mu (x_1{-}x_2{+}{\dots}
{+}x_{N'-1}{-}x_{N'})\rangle} 
\exp\Big( -\frac{N' t \|p\|^2}{2\mu^2} \Big)
& \text{ for $N'$ even}\;,
\\[0.5ex] 0 &  \text{ for $N'$ odd}\;.
\end{array}\right.
\end{align*}
Integration of 
$\mathcal{G}(t^1,\omega^1_1,\dots,\omega^1_{N_1{-}1}|
\dots | t^B,\omega^B_1,\dots,\omega^B_{N_B{-}1})$ against 
$\exp( -\frac{N_\beta t^\beta \|p^\beta\|^2}{2\mu^2})$ 
in (\ref{calG}) returns to the original function
$G_{|\under{q}^1_1 \dots \under{q}^1_{N_1}|\dots|
\under{q}^B_1 \dots \under{q}^B_{N_B}|}$, but with
\begin{enumerate}
\item for each $\beta$, all $|\under{q}^\beta_i|$ coincide
(no $\omega$-dependence),

\item 
$\frac{\sum_{i=1}^{N_\beta} |\under{q}^\beta_i|}{\sqrt{V\mu^4}}=
\frac{N_\beta}{2\mu^2} \|p\|^2$, hence 
$\frac{|\under{q}^\beta_i|}{\sqrt{V\mu^4}} \stackrel{V\mu^4\to \infty}
\longrightarrow (1{+}\mathcal{Y})q = \frac{|p|^2}{2\mu^2}$ in the limit to
the integral representation.
\end{enumerate}
We have thus proved \cite{Grosse:2013iva}:
\begin{Theorem}
  The connected $N$-point Schwinger functions of the $\phi^4_4$-model on
  extreme Moyal space $\theta\to \infty$ are given by
\begin{align}
&S_c(\mu x_1,\dots,\mu x_N)
\nonumber
\\
&= \frac{1}{64\pi^2}\!\!\!\!\!
\sum_{\di{N_1{+}\dots{+}N_B=N}{N_\beta\,\mathrm{even}}}
\sum_{\sigma \in \mathcal{S}_N} \!\!
\bigg(\prod_{\beta=1}^B \!\frac{4^{N_\beta}}{N_\beta}\!
\int_{\mathbb{R}^4} \frac{dp_\beta}{4\pi^2\mu^4}
\;e^{\mathrm{i}\big\langle
\frac{p_\beta}{\mu},\sum_{i=1}^{N_\beta} ({-}1)^{i{-}1} \mu
x_{\sigma(N_1{+}\dots{+}N_{\beta-1}+i)}\big\rangle} \bigg)
\nonumber
\\*[-0.5ex]
&\qquad\quad \times
{\mbox{\Large$G$}}_{\!\!{\underbrace{\tfrac{\|p_1\|^2}{
2\mu^2(1+\mathcal{Y})},\cdots,
\tfrac{\|p_1\|^2}{2\mu^2(1+\mathcal{Y})}}_{N_1}}\big| \dots \big|
{\underbrace{\tfrac{\|p_B\|^2}{2\mu^2(1+\mathcal{Y})},\cdots,
\tfrac{\|p_B\|^2}{2\mu^2(1+\mathcal{Y})}}_{N_B}}}\;.
\label{Schwinger-final}
\end{align}
\end{Theorem}
Some comments:

\begin{itemize} 

\item Only a restricted sector of the underlying matrix model contributes
  to position space: All strands of the same boundary component carry
  the same matrix index.

\item Schwinger functions are symmetric and invariant under the full
  Euclidean group. This comes truly surprising since $\theta\neq 0$
  breaks both translation invariance and manifest rotation invariance.
  The limit $\theta\to \infty$ was expected to make this symmetry
  violation even worse!

\item The most interesting sector is the case where  
every boundary component has $N_\beta=2$ indices. 
It is described by the ($2{+}\dots{+}2$)-point functions
$G_{\frac{\|p_1\|^2}{2\mu^2(1{+}\mathcal{Y})}
\frac{\|p_1\|^2}{2\mu^2(1{+}\mathcal{Y})}\big|\dots \big|
  \frac{\|p_B\|^2}{2\mu^2(1{+}\mathcal{Y})}
\frac{\|p_B\|^2}{2\mu^2(1 {+}\mathcal{Y})}}$.

\item This sector describes the propagation and interaction of $B$
  particles without any momentum exchange.  This is acceptable for a
  2D-model. In four dimensions, absence of momentum transfer is a sign
  of \emph{triviality}.

\item However, typical triviality proofs rely on clustering, analyticity
  in Mandelstam representation or absence of bound states. All this needs
  verification.

\end{itemize}

It is already clear that clustering is maximally violated. Looking for
instance at the  $(2{+}2)$-sector, we have 
\begin{align}
\lim_{\mu a\to  \infty}
&S^{2+2}_c(\mu x_1,\mu x_2,\mu (x_3+a),\mu (x_4+a))
\nonumber
\\
& =
\int \frac{dp\,dq}{4\pi^6\mu^4}   \;
G_{\frac{\|p\|^2}{2\mu^2(1{+}\mathcal{Y})}
\frac{\|p\|^2}{2\mu^2(1{+}\mathcal{Y})}\big|
  \frac{\|q\|^2}{2\mu^2(1{+}\mathcal{Y})}
\frac{\|q\|^2}{2\mu^2(1 {+}\mathcal{Y})}}
e^{\mathrm{i}\langle p,x_1{-}x_2\rangle+
\mathrm{i}\langle q,x_3{-}x_4\rangle}
\end{align}
independent of the distance between $\{x_1,x_2\}$ on one hand and
$\{x_3,x_4\}$ on the other hand.  Absence of clustering means that the
vacuum state (of a hypothetical continuation to a Wightman theory) is
not a pure state.  Non-pure states can be decomposed into pure states which
describe different topological sectors.

Let us give an intuitive explanation why the limit $\theta\to \infty$
of extreme noncommutativity is so close to an ordinary field theory
expected for $\theta\to 0$.
The interaction term in momentum space
\[
\frac{\lambda}{4} 
\int_{(\mathbb{R}^4)^4} \Big(\prod_{i=1}^4 \frac{dp_i}{(2\pi)^4}\Big)\,
\delta(p_1+\dots+p_4)\,\exp\Big(\mathrm{i}\sum_{i<j} 
\langle p_i,\Theta p_j\rangle\Big)\;
\prod_{i=1}^4 \hat{\phi}(p_i)
\]
leads to the Feynman rule $\lambda\exp\big(\mathrm{i}\sum_{i<j}
\langle p_i,\Theta p_j\rangle\big)$, plus momentum conservation.  For
$\theta\to \infty$, this converges to zero almost everywhere by the
Riemann-Lebesgue lemma, \emph{unless $p_i,p_j$ are linearly
  dependent}.
This case of linearly dependent momenta might be protected 
for topological reasons, and these are precisely the boundary components $B>1$
which guarantee full Lebesgue measure!

\subsection{Reflection positivity}
\label{sec:reflect}

Under conditions identified by Osterwalder-Schrader
\cite{Osterwalder:1973dx, Osterwalder:1974tc}, Schwinger functions
\cite{Schwinger:1959zz} of a Eulidean quantum field theory permit an
analytical continuation to Wightman functions \cite{Wightman:1956zz,
  Streater:1964??} of a true relativistic quantum field theory.  In
simplified terms, the reconstruction theorem of Osterwalder-Schrader
for a field theory on $\mathbb{R}^d$ says:
\begin{Theorem}[\cite{Osterwalder:1973dx, Osterwalder:1974tc}]
Assume the Schwinger functions $S(x_1,\dots, x_N)$ satisfy
\begin{enumerate} \setcounter{enumi}{-1}
\item growth conditions,

\item Euclidean covariance,

\item 
reflection positivity: for each tuple $(f_0,\dots, f_K)$ of 
test functions $f_N{\in} \mathcal{S}(\mathbb{R}^{Nd})$,
\[
\sum_{M,N=0}^K \int \! dx\, dy\; S(x_1,\dots,x_N,y_1,\dots, y_M)
\overline{f_N(x_1^r ,\dots ,x^r_N) }
f_M(y_1,\dots, y_M) \geq 0\;,
\]
where $(x^0,x^1,\dots x^{d-1})^r:=(-x^0,x^1,\dots x^{d-1})$,

\item permutation symmetry.

\end{enumerate}
Then the $S(\xi_1,\dots \xi_{N-1})\big|_{\xi_i^0>0}$, with
$\xi_i=x_i{-}x_{i+1}$, are Laplace-Fourier transforms of Wightman
functions in a relativistic quantum field theory.  If in addition the
$S(x_1,\dots, x_N)$ satisfy

\vspace*{-1ex}

\begin{enumerate} \setcounter{enumi}{3}

\item clustering
\end{enumerate}
then the Wightman functions satisfy clustering, too.
\end{Theorem}
Representation as Laplace transform in $\xi^0$ requires analyticity in
$\mathrm{Re}(\xi^0)>0$. For the Schwinger 2-point function
(\ref{Schwinger-final}), such analyticity in $\xi^0$ 
is a corollary
of analyticity of the function $a\mapsto G_{aa}$ in 
$\mathbb{C}\setminus {]{-}\infty,0]}$. We will show that analyticity
and reflection positivity boil down to \emph{Stieltjes functions},
i.e.\ functions $f:\mathbb{R}_+\to \mathbb{R}$ which have a
representation as a Stieltjes transform (see \cite{Widder:1938??})
\begin{align}
f(x)=c+ \int_0^\infty \frac{d(\rho(t))}{x+t}\;,\qquad c=f(\infty)\geq 0\;,
\end{align}
where $\rho$ is non-negative and non-decreasing.
We prove:
\begin{Proposition}
  The Schwinger function $S_c(\mu\xi)= \displaystyle
  \int_{\mathbb{R}^4} \frac{dp}{(2\pi\mu)^4} e^{\mathrm{i}p\xi}
  G_{\frac{\|p\|^2}{2\mu^2(1+\mathcal{Y})}\frac{\|p\|^2}{2\mu^2(1+\mathcal{Y})}}$
  identified in (\ref{Schwinger-final}) is the analytic continuation
  of a Wightman $2$-point function if and only if $a\mapsto G_{aa}$ is
  Stieltjes.
\end{Proposition}
\emph{Proof}. This is verified by explicit calculation. If 
$a\mapsto G_{aa}$ is Stieltjes, we have in terms of 
$\omega_{\vec{p}}(t):=\sqrt{\vec{p}^2+
2\mu^2(1+\mathcal{Y})t}$
\begin{subequations}
\begin{align}
S_c(\mu\xi)\big|_{\xi^0>0}
&=
\int_{\mathbb{R}^3} \frac{d\vec{p}}{(2\pi\mu)^3} 
\int_{-\infty}^\infty \frac{dp^0}{2\pi \mu}
e^{\mathrm{i}p^0\xi^0+\mathrm{i} \vec{p}\cdot \vec{\xi} } 
\int_0^\infty \frac{d\rho(t)}{t+ \frac{(p^0)^2+\vec{p}^2}{
2\mu^2(1+\mathcal{Y})}}
\nonumber
\\
&=2\mu(1{+}\mathcal{Y}) \!\int_{\mathbb{R}^3} \! \frac{d\vec{p}\;
e^{\mathrm{i} \vec{p}\cdot \vec{\xi} } }{(2\pi\mu)^3} 
\int_0^\infty \!\! \frac{d\rho(t)}{2\omega_{\vec{p}}(t)} 
\int_{-\infty}^\infty \!\frac{dp^0}{2\pi \mathrm{i}} 
\Big(\frac{e^{\mathrm{i}p^0\xi^0}}{p^0{-}\mathrm{i}\omega_{\vec{p}}(t)}
-
\frac{e^{\mathrm{i}p^0\xi^0}}{p^0{+}\mathrm{i}\omega_{\vec{p}}(t)}\Big)
\nonumber
\\
&=2\mu(1{+}\mathcal{Y}) 
\int_0^\infty \!\! \frac{d\rho(t)}{2\omega_{\vec{p}}(t)} 
\!\int_{\mathbb{R}^3} \!\! \frac{d\vec{p}\;
e^{-\xi^0\omega_{\vec{p}}(t) + \mathrm{i} \vec{p}\cdot \vec{\xi} } }{(2\pi\mu)^3} 
\nonumber
\\
&=\int_0^\infty \frac{2(1+\mathcal{Y}) \,d\rho(t)}{\mu^4} 
\int_0^\infty \!\!dq^0\int_{\mathbb{R}^3}\!\! 
d\vec{q} \; \hat{W}_t(q) 
e^{-q^0\xi^0+\mathrm{i}\vec{q}\cdot \vec{x}}\;,
\\
\hat{W}_t(q) &:=\frac{\theta(q^0)}{(2\pi)^3} \delta\Big(\frac{(q^0)^2
-\vec{q}^2-2\mu^2(1{+}\mathcal{Y})t}{\mu^2}\Big)\;.
\end{align}
\end{subequations}
The step from the second to third line is the residue theorem.  We
observe that $\hat{W}_t(q)$ is precisely the K\"all\'en-Lehmann
spectral representation \cite{Kallen:1952zz, Lehmann:1954xi} of a
Wightman 2-point function.  \hfill $\square$%

\bigskip

Remarkably,  the Stieltjes property can be tested by purely
real conditions:
\begin{Theorem}[Widder \cite{Widder:1938??}]
A function $f:\mathbb{R}_+{\to} \mathbb{R}$ is Stieltjes iff it is smooth,
non-negative and satisfies $L_{k,t}[f(\bullet)] \geq 0$, where
\[
L_{k,t}[f(\bullet)]
:= \frac{(-t)^{k-1}}{c_k}
\frac{d^{2k-1}}{dt^{2k-1}} \big( 
t^k f(t)\big)\;,\qquad c_1=1,\;c_{k>1}=k!(k{-}2)! \;.
\]
In that case, the measure is recovered by
$\rho'(t)=\lim_{k\to \infty} L_{k,t}[f(\bullet)]$ (weakly and almost
everywhere). 
\end{Theorem}
The perturbatively established anomalous dimension $\eta=-2\lambda$
implies that $a\mapsto G_{aa}$ cannot be Stieltjes for $\lambda>0$.
The restriction to negative coupling constant is reminiscent of the
planar wrong-sign $\lambda\phi^4_4$-model \cite{'tHooft:1982cx,
  Rivasseau:1983jj}. Recall that our matrix model also reduces to the
planar sector, but as result of the infinite volume limit and not by
hand. We nonetheless keep a non-trivial topology in form of $B\geq 1$
boundary components. Moreover, we have an exact solution for
$S(x_1,\dots,x_N)$, not only an existence proof.

\begin{figure}[h!]
\begin{picture}(150,90)
  \put(0,45){\includegraphics[width=7cm,
viewport=0 0 288 182]{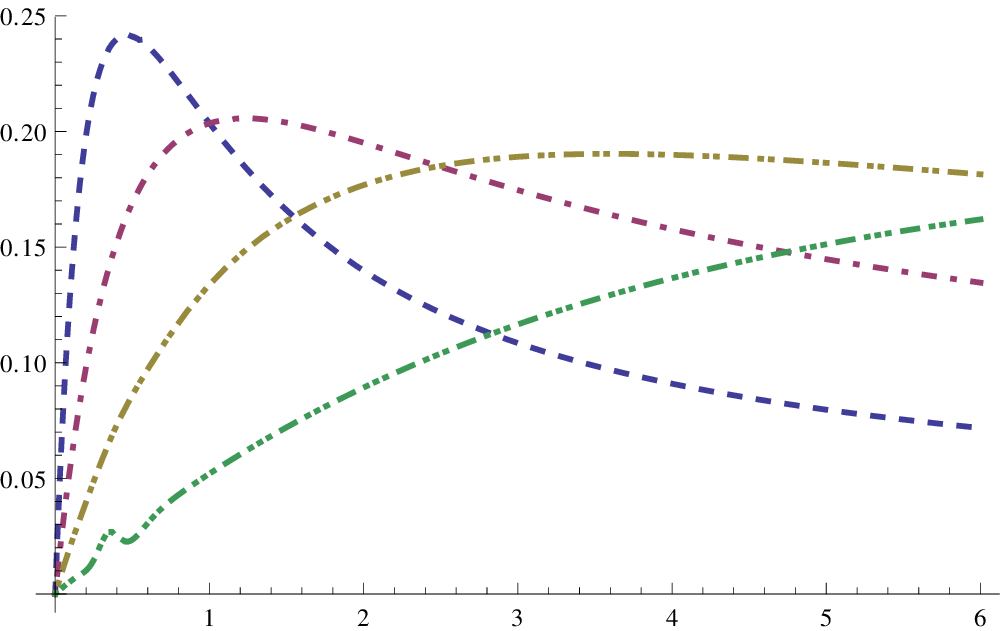}}
  \put(56,57){\mbox{\scriptsize$\lambda{=}{-}0.350$}}
  \put(17,82){\mbox{\scriptsize$\lambda{=}{-}0.382$}}
  \put(44,80){\mbox{\scriptsize$\lambda{=}{-}0.398$}}
  \put(11,51){\mbox{\scriptsize$\lambda{=}{-}0.414$}}
  \put(32,54){\fbox{\mbox{\small$L_2$}}}
  \put(75,45){\includegraphics[width=7cm,
viewport=0 0 288 169]{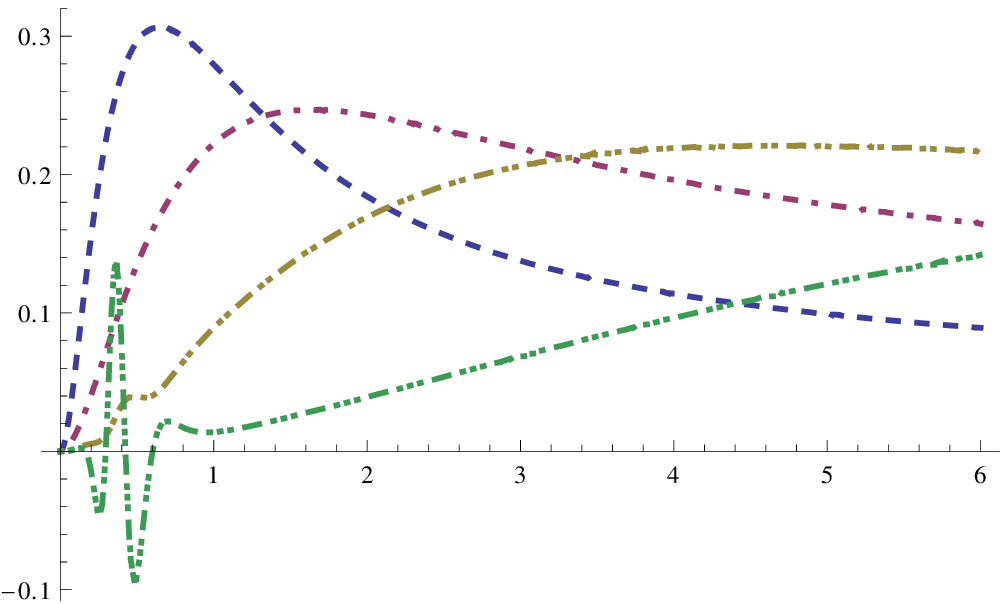}}
  \put(89,84){\mbox{\scriptsize${-}0.350$}}
  \put(100,80){\mbox{\scriptsize${-}0.382$}}
  \put(130,78){\mbox{\scriptsize${-}0.398$}}
  \put(85,48){\mbox{\scriptsize${-}0.414$}}
  \put(125,48){\fbox{\mbox{\small$L_3$}}}
  \put(0,0){\includegraphics[width=7cm,
viewport=0 0 288 171]{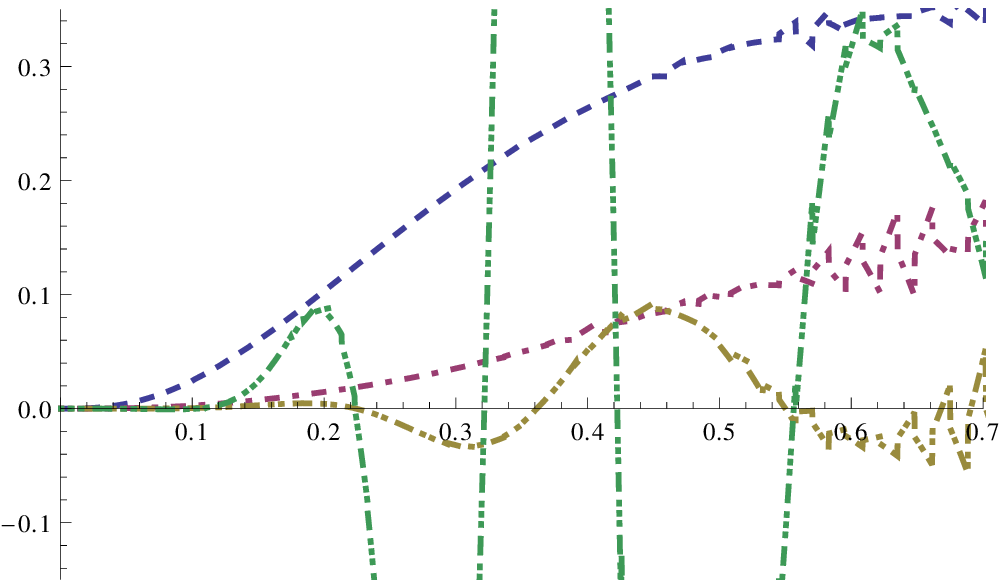}}
  \put(19,25){\mbox{\scriptsize${-}0.350$}}
  \put(59,18){\mbox{\scriptsize${-}0.382$}}
  \put(60,5){\mbox{\scriptsize${-}0.398$}}
  \put(16,2){\mbox{\scriptsize${-}0.414$}}
  \put(10,30){\fbox{\mbox{\small$L_4$}}}
\put(75,20){\parbox{7cm}{\begin{itemize}
    \item based on interpolation of discrete data, noisy for $k\geq 4$
      
    \item Stieltjes property clearly violated for $\lambda<\lambda_c$ 

    \end{itemize}
  }}
\end{picture}
\caption{Widder's criteria 
$L_{k,a}[G_{\bullet\bullet}]:=
\frac{(-a)^{k-1}}{k!(k-2)!}\frac{d^{2k-1}}{da^{2k-1}}(a^k G_{aa})\geq 0
$ for $\lambda \approx \lambda_c$.\label{fig3}}
\end{figure}
\begin{figure}[h!]
\begin{picture}(150,90)
\put(0,47){\includegraphics[width=7cm,
          viewport=0 0 288 178]{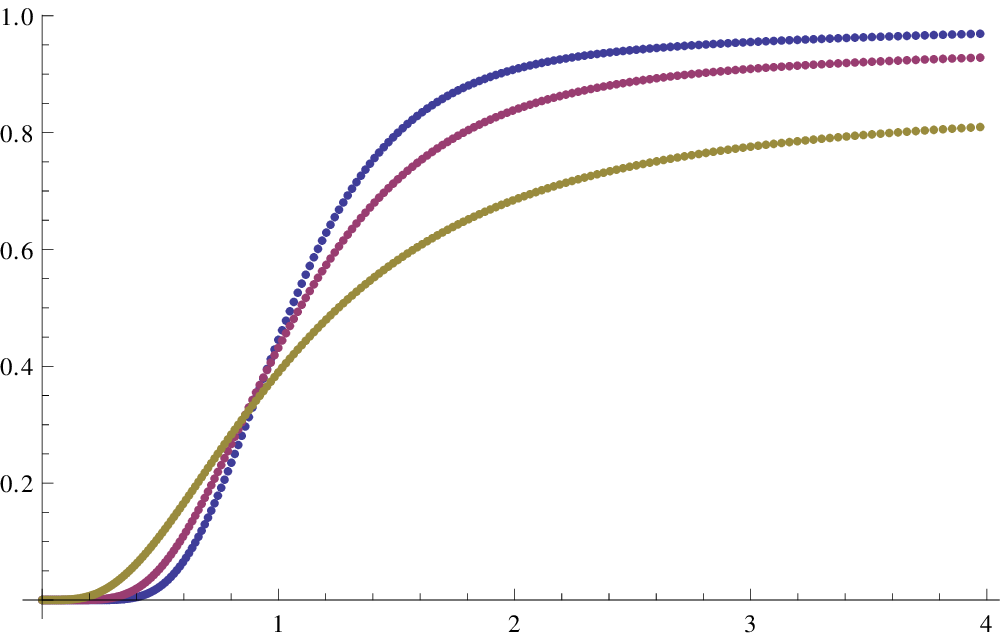}}
  \put(35,60){\fbox{\mbox{\small$\lambda=-0.032$}}}
  \put(54,78.5){\mbox{\small$\tilde{\rho}_5$}}
  \put(36,81){\mbox{\small$\tilde{\rho}_{10}$}}
  \put(21,82){\mbox{\small$\tilde{\rho}_{16}$}}
\put(75,47){\includegraphics[width=7cm,
          viewport=0 0 288 178]{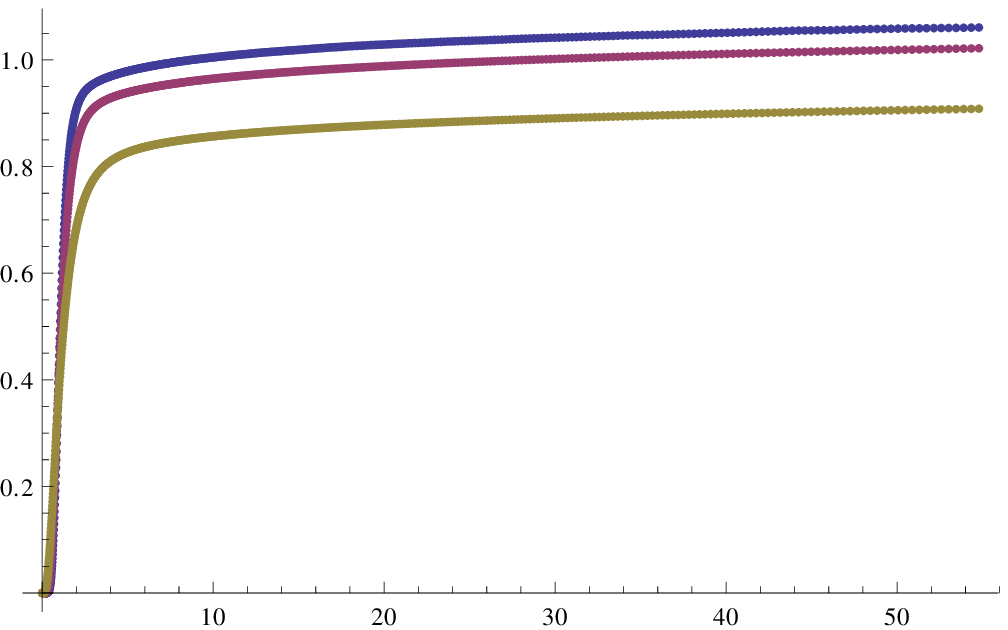}}
  \put(110,70){\fbox{\mbox{\small$\lambda=-0.032$}}}
  \put(130,80){\mbox{\small$\tilde{\rho}_5$}}
  \put(90,89){\mbox{\small$\tilde{\rho}_{16}$}}
\put(0,0){\includegraphics[width=7cm,
          viewport=0 0 288 178]{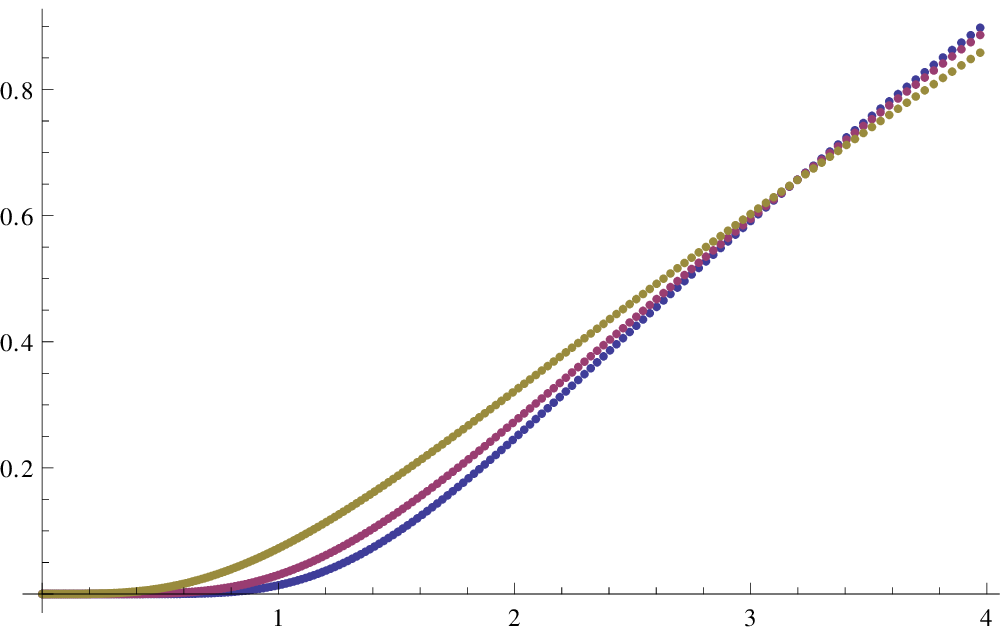}}
  \put(23,12){\mbox{\small$\tilde{\rho}_5$}}
  \put(31,7){\mbox{\small$\tilde{\rho}_{16}$}}
  \put(8,30){\fbox{\mbox{\small$\lambda\pi=-0.366$}}}
\put(75,0){\includegraphics[width=7cm,
          viewport=0 0 288 178]{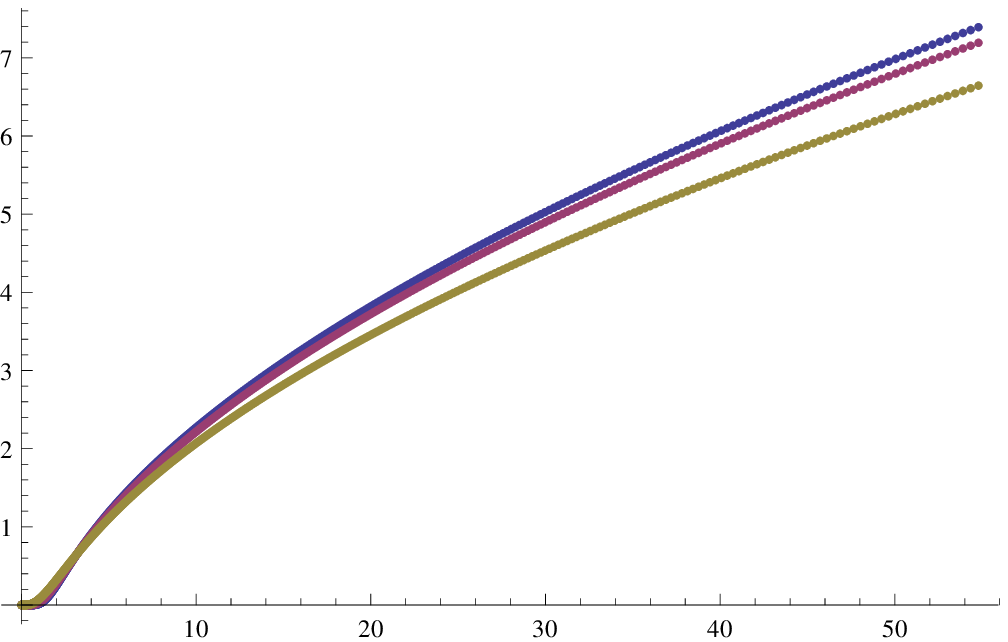}}
  \put(82,34){\fbox{\mbox{\small$\lambda=-0.366$}}}
  \put(124,29){\mbox{\small$\tilde{\rho}_5$}}
  \put(124,39){\mbox{\small$\tilde{\rho}_{16}$}}
\thicklines
\put(100,17){\mbox{\small critical behaviour:}}
\put(93,9){\mbox{\small 
$ \tilde{\rho}(\mu^2){=}\left\{ \!\!\!\begin{array}{cl} 0 & \mu \leq m \\
(\mu^2{-}m^2)^{-\alpha} & \mu\geq m
\end{array}\right.$}}
\put(108,60){\mbox{\small close to step function}}
\put(108,55){\mbox{\small expected for $\lambda=0$}}
\end{picture}
\caption{$\tilde{\rho}_k(m^2)=\int_0^{m^2} dt\;L_{k,t}[G_{\bullet 0}]$
as approximation for the mass density of $a\mapsto G_{aa}$.
In each row, the left picture is zoomed into small $\mu$, showing
evidence for a mass gap. The right pictures show the global
behaviour, close to a step function for $\lambda\nearrow 0$, close to
criticality for $\lambda\searrow \lambda_c$.
\label{fig4}}
\end{figure}

Whether or not $a\mapsto G_{aa}$ is a Stieltjes function for
$\lambda<0$ is a highly interesting question. A first idea can be
obtained by computer simulations, see sec.~\ref{sec:computer}. We show
in Figure~\ref{fig3} interpolation results for $\lambda$ near the
critical coupling constant.
We find clear evidence that $a\mapsto G_{aa}$ is not a Stieltjes
function for $\lambda<\lambda_c$, where $\lambda_c\approx {-}0.396$
locates the discontinuity of $\mathcal{Y}'(\lambda)$.  For $\lambda\in
[\lambda_c,0]$ the results are not conclusive (as $k$ is too small).
Since $G_{aa}$ and $G_{a0}$ show a very similar behaviour (see e.g.\
Fig.~\ref{fig2}), the functions $L_{k,t}[G_{\bullet 0}]$ (which are
easy to compute) give some indication about $L_{k,t}[G_{\bullet
  \bullet}]$ (which we are interested in). From (\ref{master}) one can
prove the following identity \cite{Grosse:2014??}:
\begin{align}
\frac{(\log G_{a0})^{(\ell)}}{(\ell-1)!}
=  \frac{(-1)^{\ell}}{(1{+}a)^\ell}
+ (-1)^{\ell} \,\mathrm{sign}(\lambda) \,
\mathcal{H}_0^{\!\Lambda}\Big[
\sin\big( \ell \tau_a(\bullet)\big)
\Big(\frac{\sin \tau_a(\bullet)}{|\lambda|\pi \bullet}\Big)^{\!\ell}
\Big]\;.
\end{align}
The resulting integrated `mass densities'
$\tilde{\rho}_k(m^2)=\int_0^{m^2} dt\;L_{k,t}[G_{\bullet 0}]$ are
shown in Figure~\ref{fig4}.
We find clear evidence for a mass gap, $\lim_{k\to
  \infty}\tilde{\rho}_k(\mu^2)=0$ for $0\leq \mu^2 \leq m^2$. For
$\lambda\nearrow 0$ the integrated mass density approaches (as
expected) a step function, whereas for $\lambda\searrow \lambda_c$ we
notice a power-law behaviour typical for critical phenomena. In
particular, for $\lambda_c < \lambda<0$ there is no further gap in the
support of $\tilde{\rho}'$, which signals scattering right away from
$m^2$ (not only from the two-particle threshold on). We interpret this
as scattering of a massive particle with an infrared cloud.
This scattering would be a remnant of the underlying non-trivial
matrix model before the projection to diagonal matrices.

\subsection{Summary}

We have shown that the $\phi^4_4$-model on noncommutative Moyal space,
considered in the limit $\theta\to \infty$ of extreme
noncommutativity, is an exactly solvable and non-trivial matrix model.
Euclidean symmetry is violated in the beginning, but we identified a
limit which projects to diagonal matrices where Euclidean symmetry is
restored.  One would not expect that such a brutal projection can
respect any quantum field theory axioms. Surprisingly, the first
consistency checks, positivity of the lowest Widder criteria
$L_{k,t}[G_{\bullet\bullet}]$, are passed for the only interesting
interval $[\lambda_c,0]$ of the coupling constant!

If these miracles continue and all Osterwalder-Schrader axioms (except
for clustering) hold, we would get a relativistic quantum field theory
in four dimensions. This theory is somewhat strange as `particles'
keep their momenta in interaction processes. Nevertheless, the theory
is not completely trivial.  We find scattering remnants from the
noncommutative geometrical (i.e.\ matricial) substructure. Only the
external matrix indices are put `on-shell', internally all degrees of
freedom contribute. 

We have seen that clustering is maximally violated.  The interaction
is insensitive to positions in different boundary components. 
In particular, `particles' are never asymptotically free.

\begin{appendix}

\section{Schwinger-Dyson equations for $B=2$}

We find for the $(1{+}1)$- and $(2{+}2)$-point
functions 
\begin{subequations}
\label{G1+1}
\begin{align}
G_{|a|c|}
&= 
- \frac{\lambda}{E_a{+}E_a} \bigg(
\frac{1}{V}\sum_{p\in I} \Big(
 G_{|ap|} G_{|a|c|}
- \frac{G_{|p|c|}{-}G_{|a|c|}}{E_p{-}E_a}\Big)
- \frac{  G_{|cc|}{-} G_{|ac|}}{E_c{-}E_a}\bigg) && \bigg\}\!\!\!\! 
\label{G11-a}
\\[-.3ex]
& 
- \frac{\lambda}{V^2(E_a{+}E_a)}\Big( 
3G_{|a|a|}G_{|a|c|} + G_{|a|cac|}+ G_{|c|aaa|}
+ \frac{1}{V}\sum_{n\in I} G_{|a|c|an|}\Big)&& \bigg\}\!\!\!\!  
\label{G11-b}
\\[-.3ex]
&
- \frac{\lambda}{V^4(E_a{+}E_a)} G_{|a|a|a|c|}\;, && \bigg\}\!\!\!\!  
\label{G11-c}
\end{align}
\end{subequations}

\vspace*{-2.8ex}

\begin{subequations}
\label{G2+2}
\begin{align}
&G_{|ab|cd|}
\nonumber
\\
& \begin{array}[c]{@{}l@{}}
\displaystyle 
= - \frac{\lambda}{E_a{+}E_b} \bigg(
\frac{1}{V}\sum_{p\in I} \Big(
\big(G_{|ap|} G_{|ab|cd|}
{+}G_{|ab|} G_{|ap|cd|}\big)
- \frac{G_{|pb|cd|}{-}G_{|ab|cd|}}{E_p-E_a}\Big)
\\
\displaystyle 
\quad\quad
+G_{|ab|}\big(G_{|cacd|}{+}G_{|dadc|}\big)
-\frac{  G_{|cbcd|}{-} G_{|cbad|}}{E_c-E_a}
-\frac{  G_{|dbdc|}{-} G_{|dbac|}}{E_d-E_a}
\bigg)
\end{array} && \left.\rule{0mm}{12mm}\right\}
\label{G22-a}
\\[-.3ex]
&
\begin{array}[c]{@{}l@{}}
\displaystyle 
- \frac{\lambda}{V^2(E_a+E_b)}\Big( 
G_{|a|a|}G_{|ab|cd|} +G_{ab} G_{|a|a|cd|}
+ \frac{1}{V}\sum_{n\in I} G_{|an|ab|cd|}
\\*
\quad\quad
\displaystyle+ G_{|cd|aaab|}{+} G_{|cd|baba|}
{+} G_{|ab|cacd|}{+} G_{|ab|cddad|}
- \frac{  G_{|b|a|cd|}{-} G_{|b|b|cd|}}{E_b-E_a}\Big)
\end{array}
&& \left.\rule{0mm}{12mm}\right\}
\label{G22-b}
\\*[-.3ex]
& 
- \frac{\lambda}{V^4(E_a+E_b)} G_{|a|a|ab|cd|}\;.
&& \left.\rule{0mm}{6mm}\right\}
\label{G22-c}
\end{align}
\end{subequations}
These are basic functions which are not simplified by reality.  As
before, (\ref{G11-a}) and (\ref{G22-a}) preserve the genus, whereas 
$g\mapsto
g{+}1$ in (\ref{G11-b})+(\ref{G22-b}) and $g\mapsto g{+}2$ in
(\ref{G11-c})+(\ref{G22-c}).  The higher $(N_1{+}N_2)$-point functions
with one $N_i\geq 3$ simplify by reality to universal recursion
formulae. For $N_i$ odd we have
\begin{align}
&
G_{|b_0\dots b_{2l}|c_1\dots c_{N-2l-1}|}
\nonumber
\\
&=
-\lambda\! \sum_{k=1}^{N-2l-1}
\frac{G_{|c_1\dots c_{k-1} b_0 b_1\dots b_{2l}
c_k c_{k+1}\dots c_{N-2l-1}|}
{-}G_{|c_1\dots c_{k-1} c_k b_1\dots b_{2l} b_0
c_{k+1}\dots c_{N-2l-1}|}
}{(E_{b_1}-E_{b_{2l}})(E_{b_0}-E_{c_k})}
\nonumber
\\
&
-\lambda \! \sum_{j=1}^{l}
\frac{G_{|b_0b_1\dots b_{2j-2}|c_1\dots c_{N-2l-1}|}
G_{|b_{2j-1}b_{2j}\dots  b_{2l}|}
{-}G_{|b_{2j-1}b_1\dots b_{2j-2}|c_1\dots c_{N-2l-1}|}
G_{|b_0b_{2j}\dots b_{2l}|}
}{(E_{b_1}-E_{b_{2l}}) (E_{b_0}-E_{b_{2j-1}})}
\nonumber
\\
&
-\lambda \! \sum_{j=1}^{l}
\frac{G_{|b_0b_1\dots b_{2j-1}|}
G_{|b_{2j}b_{2j+1}\dots  b_{2l}|c_1\dots c_{N-2l-1}|}
{-}G_{|b_{2j}b_1\dots b_{2j-1}|}
G_{|b_0b_{2j+1}\dots b_{2l}|c_1\dots  c_{N-2l-1}|} }{
(E_{b_1}-E_{b_{2l}})(E_{b_0}-E_{b_{2j}})}
\nonumber
\\
&
-
\frac{\lambda}{V^2} \sum_{k=1}^{2l}
\frac{G_{|b_0b_1\dots b_{k-1}|b_kb_{k+1}\dots b_{2l}|
c_1\dots c_{N-2l-1}|}{-}
G_{|b_kb_1\dots b_{k-1}|b_0 b_{k+1}\dots b_{2l}|c_1\dots
  c_{N-2l-1}|}}{(E_{b_1}-E_{b_{2l}})(E_{b_0}-E_{b_k})}\;.
\label{GN-B=2-odd}
\end{align}
The last line increases the genus and is absent in $G^{(0)}_{|b_0b_1\dots
  b_{2l}|c_1\dots c_{N-2l-1}|}$. For $N_i$ even one finds
\begin{align}
& 
G_{|a b_1\dots b_{2l-1}|c_1\dots c_{N-2l}|}
\nonumber
\\
&= -\lambda 
\sum_{j=1}^{l-1}
\frac{
G_{|b_1\dots b_{2j-1}a|c_1\dots c_{N-2l}|}
G_{|b_{2j}b_{2j+1}\dots b_{2l-1}|}
{-}
G_{|b_1\dots b_{2j-1}b_{2j}|c_1\dots c_{N-2l}|}
G_{|ab_{2j+1}\dots b_{2l-1}|}
}{(E_{b_1}-E_{b_{2l-1}})(E_a-E_{b_{2j}})}
\nonumber
\\
&
-\lambda 
\sum_{j=1}^{l-1}
\frac{
G_{|b_1\dots b_{2j-1}a|}
G_{|b_{2j}b_{2j+1}\dots b_{2l-1}|c_1\dots c_{N-2l}|}
{-}
G_{|b_1\dots b_{2j-1}b_{2j}|}
G_{|ab_{2j+1}\dots b_{2l-1}|c_1\dots c_{N-2l}|}
}{(E_{b_1}-E_{b_{2l-1}})(E_a-E_{b_{2j}})}
\nonumber
\\
&
-\lambda \sum_{k=1}^{N-2l}
\frac{
G_{|c_1\dots c_{k-1}ab_1\dots b_{2l-1}c_k c_{k+1}\dots c_{N-2l}|}
{-}G_{|c_1\dots c_{k-1}c_kb_1\dots b_{2l-1}a c_{k+1}\dots c_{N-2l}|}
}{(E_{b_1}-E_{b_{2l-1}})(E_a-E_{c_k})}
\nonumber
\\
&
-\frac{\lambda}{V^2} \sum_{k=1}^{2l-1}\frac{
G_{|b_1\dots b_{k-1}a|b_kb_{k+1}\dots b_{2l-1}|c_1\dots c_{N-2l}|}{-}
G_{|b_1\dots b_{k-1}b_k|ab_{k+1}\dots b_{2l-1}|c_1\dots c_{N-2l}|}
}{(E_{b_1}-E_{b_{2l-1}})(E_a-E_{b_k})}\;.
\label{GN-B=2-even}
\end{align}
Again, the last line increases the genus and is absent 
in $G^{(0)}_{|b_0b_1\dots b_{2l-1}|c_1\dots C_{N-2l}|}$.

\end{appendix}

\section*{Acknowledgements}

RW would like to cordially thank Keiichi R.\ Ito for the invitation to
the RIMS symposium ``Applications of RG Methods in Mathematical
Sciences'' and for the immense help and hospitality during 
this visit to Kyoto in September 2013.

In addition to the presentation at the RIMS symposium, this
contribution is based on a lecture series which both of us gave in
November 2013 in G\"ottingen.  We would like to thank Dorothea Bahns
for this arrangement and for hospitality during this series.

\end{document}